\documentclass[superscriptaddress,aps,prb,amsmath,amssymb,twocolumn,letterpaper,
         footinbib,bibnotes,showpacs,reprint,balancelastpage,raggedbottom,longbibliography,
         citeautoscript,floatfix]{revtex4-2}

\usepackage[usenames,dvipsnames,table]{xcolor}
\usepackage[%
  breaklinks,             %
  bookmarks = false, %
  pdfpagemode   = UseNone,%
  pdfstartview  = FitH,     %
  pdfstartpage  = 1,      %
  colorlinks    = true,   %
]{hyperref}
\hypersetup{
    linkcolor=RubineRed,          
    citecolor=RoyalBlue,        
    filecolor=Mulberry,      
    urlcolor=RoyalBlue           
}
\usepackage{amssymb}
\usepackage{graphicx}
\usepackage{float}
\usepackage{enumerate}
\usepackage{microtype}
\usepackage{xspace}
\usepackage{nicefrac}
\usepackage{soul}  

\usepackage[utf8]{inputenc}
\DeclareUnicodeCharacter{2212}{-}

\usepackage{xr}


\begin{document}

\author{Yongjin Shin}
\affiliation{Pritzker School of Molecular Engineering, University of Chicago, Chicago, Illinois 60637, United States}
\author{Giulia Galli}
\email{gagalli@uchicago.edu}
\affiliation{Pritzker School of Molecular Engineering, University of Chicago, Chicago, Illinois 60637, United States}
\affiliation{Department of Chemistry, University of Chicago, Chicago, Illinois 60637, United States}
\affiliation{Center for Molecular Engineering and Materials Science Division, Argonne National Laboratory, Lemont, Illinois 60439, United States}

\def\degree{$^\circ$\xspace}
\newcommand\Tstrut{\rule{0pt}{2.6ex}}         
\newcommand\Bstrut{\rule[-0.9ex]{0pt}{0pt}}   
\newcommand\RAFO{$R_{1/3}A_{2/3}$FeO$_{2.67}$\xspace}
\newcommand\NCFO{Nd$_{1/3}$Ca$_{2/3}$FeO$_{2.67}$\xspace}
\newcommand\LSFO{La$_{1/3}$Sr$_{2/3}$FeO$_{2.67}$\xspace}
\newcommand\SFO{SrFeO$_{2.5}$\xspace}
\newcommand\CFO{CaFeO$_{2.5}$\xspace}
\newcommand\BFO{BaFeO$_{2.5}$\xspace}
\newcommand\AFO{$A$FeO$_{2.5}$\xspace}
\newcommand\Ccm{$\mu\mathrm{C/cm}^2$\xspace}

\title{Tunable ferroelectricity in oxygen-deficient perovskites}


\begin{abstract}
Using first-principles calculations, we predict that tunable ferroelectricity can be realized in oxide perovskites with the Grenier structure and ordered oxygen vacancies. Specifically, we show that \RAFO solids (where $R$ is a rare-earth ion and $A$ an alkaline-earth cation) exhibit stable polar phases, with a spontaneous polarization tunable by an appropriate choice of  $R$ and $A$. We find that larger cations combined with small $R$ elements lead to a maximum in the polarization and to a minimum in the energy barriers required to switch the sign of the polarization.
Ferroelectricity arises from cooperative distortions of octahedral and tetrahedral units, where a combination of rotational and sliding modes controls the emergence of polarization within  three-dimensional connected layers.
Our results indicate that polar Grenier phases of oxide perovskites are promising materials for microelectronic applications and, in general, for the study of phenomena emerging from breaking inversion symmetry in solids.
\end{abstract}

\maketitle

\section{Introduction}

Ferroelectric materials have found many interesting applications in electronic and memory devices, and understanding and engineering their properties is a topic of great interest in condensed matter physics and materials science \cite{Ramesh/Schlom2019,Sicron1994,Stern2004,Khan2014,Lines1977,Iniguez2019}. 
Ferroelectricity can be realized in polar solids without inversion symmetry, for example by relying on second-order Jahn-Teller distortions occurring in systems composed of elements with $d^0$ electronic configurations \cite{Bersuker2013}, or on the presence of \textit{lone pair}  ns$^2$ configurations e.g., in heavy elements \cite{Sicron1994}.

Oxides represent an interesting class of materials where ferroelectricity has been realized. While there are only few transition metal oxides with partially filled $d$-orbitals \cite{Spaldin2000,Spaldin_review2019}, some layered oxide perovskites structures exhibit a ferroelectric behavior due to  the geometric arrangements of layers \cite{Bousquet2008,Fennie/Rabe2005,HIF:Benedek2011,Benedek2015}. An emerging route to stabilize a polar phase in oxides is the  utilization of symmetry breaking in oxygen-deficient perovskites ($AB$O$_{3-\delta}$), for example brownmillerites where oxygen vacancies are perfectly ordered \cite{Rondinelli/May2019,Anderson1993,Chaturvedi/Leighton2021,Leighton2020}.

\begin{figure}[b]
\centering
\includegraphics[width=0.95\columnwidth]{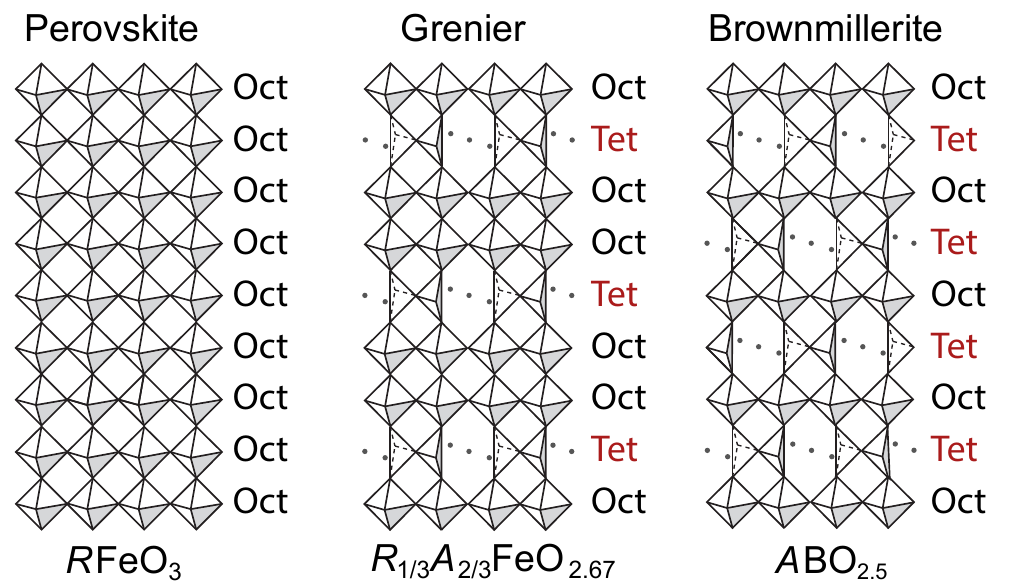}
\caption{
\textbf{Structures of oxygen-deficient perovskites.}
Schematic illustration of the $R$FeO$_3$ perovskite (left), \RAFO Grenier (middle), and \AFO brownmillerite structure (right).
%
$R$-ions and $A$-cations are omitted for clarity, and vacancy sites are drawn as gray dots.
Note that the stacking sequence of the 
Grenier phase is  OOTOOT\dots, with
two octahedral (O) layers sandwiched between tetrahedral (T)
layers, and that of the browmillerite phase is OTOTOT\dots.
}
\label{fig:schematic}
\end{figure}

\begin{figure}[t]
\centering
\includegraphics[width=0.90\columnwidth]{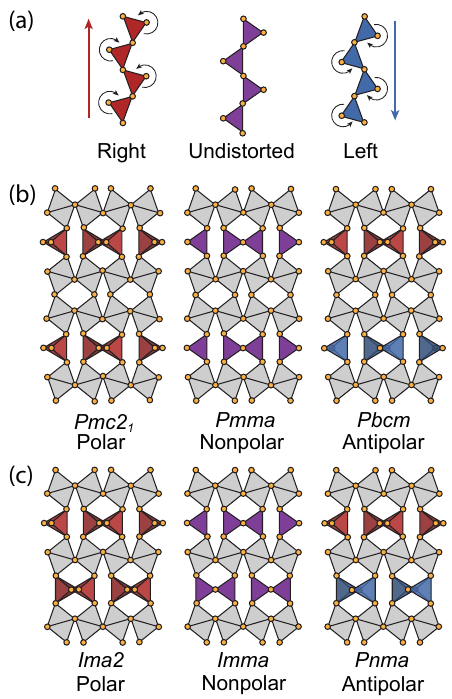}
\caption{
\textbf{Arrangements of tetrahedral chains and corresponding distortion patterns in oxygen-deficient phases.}
Twisting of tetrahedral chains (a) in oxygen-deficient perovskites with ordered oxygen vacancies.
Symmetry breaking in (b) brownmillerite and (c) Grenier phases due to the
twisting of tetrahedral chains. We omitted $A$-cations and $R$-ions for clarity,
and oxygen atoms are depicted as yellow circles. Note the
difference in space group, given below each structure, depending
on the twisting of the chains. In high symmetry, nonpolar
phases, the chains are in undistorted positions. In  polar phases, all tetrahedral chains are aligned, while  in antipolar phases two nearest tetrahedral layers exhibit opposite
chain twisting. Other distortion patterns induced
by chain twisting and observed in large unit cells are discussed
in the Supplementary Information (SI).
}
\label{fig:symmtery}
\end{figure}

In oxides with the $AB$O$_3$ perovskite structure shown in \autoref{fig:schematic},
oxygen vacancies with long-range order are formed along specific crystallographic directions, leading to the transformation of $B$O$_6$ octahedral units into $B$O$_4$ tetrahedral chains. Depending on how these chains  are `twisted' (see \autoref{fig:symmtery})  polar structures may be formed; however, in many instances such structures \cite{Tian2018,Young2015,Parsons2009,Marik/LeBreton2019} are  metastable or only marginally stable \cite{Young2017,Tian2018,Kang/Choi2019,Arras/Bellaiche2023}, thus limiting their use in devices.

One promising oxygen-deficient perovskite is the so-called Grenier phase, whose oxygen deficiency ($\delta$ = $\nicefrac{1}{3}$) is smaller than that of brownmillerites ($\delta$ = 0.5) \cite{Grenier1981,Yao/Dijken2017,Luo/Hayward2013}. 
Grenier and brownmillerite phases differ by the stacking and periodicity  of tetrahedral chains (\autoref{fig:schematic}) formed by oxygen vacancies.  
While brownmillerite solids have been extensively investigated for a variety of applications, e.g. as possible constituents of  neuromorphic devices or as ionic conductors \cite{Zhang/Galli2020,Lu2017,Shaula2006,Orera2010,PuYu2022}, Grenier phases are much less studied \cite{Karki2020,May2014,May/Spanier2015,Battle1990,May2016,Shimakawa2017,Shimakawa2022,Hudspeth2009}.
However, recent experiments observed a spontaneous polarization coexisting with antiferromagnetism in  R$_{1.2}$Ba$_{1.2}$Ca$_{0.6}$Fe$_3$O$_8$ (R= Gd, Tb) \cite{Shimakawa/Martin2019}, pointing at the possibility of realizing  stable polar Grenier phases and hence interesting ferroelectric materials, which could be used, for example, as neuromorphic devices \cite{Spaldin_review2019,Khan2020}.

In this work, using first principles calculations we predict that tunable ferroelectricity can be realized  in oxygen-deficient \RAFO (=$RA_2$Fe$_3$O$_8$) perovskites with ordered oxygen vacancies.
In particular we show that polar Grenier and antipolar brownmillerite phases may be realized, with 
the polarization determined by cooperative distortions that can be tuned by the choice of the cations.
Specifically, larger $A$ alkali-earth cations, e.g., Ba and smaller $R$ rare-earth elements, e.g.  Tb,  lead to a maximum in the spontaneous polarization of the solid and to a minimum in the energy barriers required to induce a  switch in the sign of the polarization.
%

\section{Results}

Using first principles calculations based on Density Functional Theory (DFT), we investigated the properties of \RAFO solids as a function of the rare-earth element $R$ and cation $A$. We considered $A$ =  Ca, Sr, and Ba and lanthanides from La to Tm, which are known to form $R$FeO$_3$ perovskite structures \cite{Zhang/Ansermet2021,Aparnadevi2016,Slawinski2005,Ritter2022,Ritter2021,Prakash2019,Cao2014,Bertaut1967}.
We excluded Yb and Lu from our study as  those elements were found to lead to metallic Grenier structures.

\begin{table}
\caption{\label{table:energetics} 
Energy differences ($\Delta$E, meV/f.u.) referred to
the ground state energies for prototypical oxygen-deficient perovskites.
}
\centering
\begin{ruledtabular}
\begin{tabular}{l|cc}
System   &  \NCFO       & \CFO \Bstrut \\
\hline\Tstrut
Structure type & Grenier & Brownmillerite\\
$\Delta E$ (neutral)    & 263.1 & 330.1 \\
$\Delta E$ (polar)      & 0 & 22.24  \\
$\Delta E$ (antipolar) & 38.01 & 0 \\
\end{tabular}
\end{ruledtabular}
\end{table}

\begin{figure}[ht]
\centering
\includegraphics[width=0.95\columnwidth]{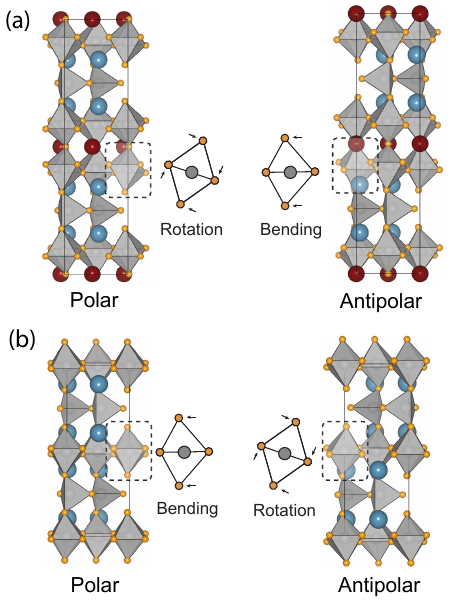}
\caption{
\textbf{Distortion patterns in polar and antipolar phases.}
Octahedral distortions in polar and antipolar phases of (a) Grenier (\RAFO) and (b) brownmillerite (\AFO) structures. Red and blue spheres indicate $R$-ions and $A$-cations, respectively. Oxygen atoms are represented as yellow circles.
}
\label{fig:distortion}
\end{figure}

\subsection{Ferroelectricity from cooperative distortions}

We start by considering  \NCFO and \CFO as representative compounds with the Grenier and brownmillerite structure, respectively, because these materials have been investigated in several experiments \cite{May2014,Hudspeth2009,Shimakawa2017,Young2017}.
In these systems, the oxidation state of Fe is 3+; Fe has a high-spin $d^5$ electronic configuration, leading to antiferromagnetic (AFM) 
Fe--Fe coupling, based on 
Goodenough-Kanamori-Anderson (GKA) superexchange rules \cite{Goodenough1963,Kanamori1959,Anderson1959}.
Hence in our calculations we constrained \NCFO and \CFO to have G-type AFM order in all brownmillerite and Grenier phases, consistent with the findings of previous studies \cite{Shimakawa/Martin2019,Young2015,Hudspeth2009,Shin/Rondinelli2020,Shin/Rondinelli2022}.
(See SI for details on magnetic ordering).

\begin{figure*}[t]
\centering
\includegraphics[width=0.85\textwidth]{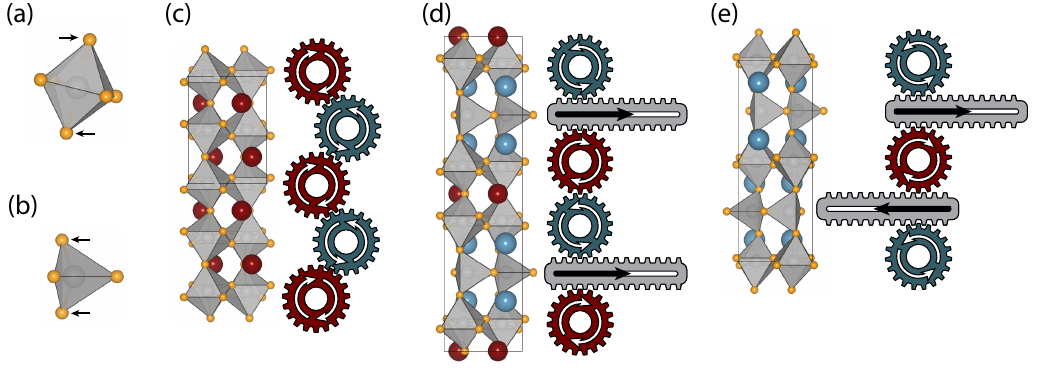}
\caption{
\textbf{Cooperative distortions and gear motions.}
Cooperative distortions of polyhedral units in perovskite, brownmillerite, and Grenier phase:
(a) Rotation of octahedra and (b) twisting of tetrahedral chains.
Stable distortion patterns in (c) $R$FeO$_3$ perovskite, (d) \RAFO, and (e) $A$FeO$_{2.5}$, where octahedral and tetrahedral layers are shown as rotational and sliding gears, respectively.
For brownmillerite, there is a single rotational gear (octahedral layer) between sliding gears (tetrahedral layers) and the two nearest sliding gears move in opposite direction.
In Grenier phases, the additional rotational gear (arising from the different  octahedral layer stacking) leads to an opposite rotation, and the sliding gears move in the same direction.
}
\label{fig:gear}
\end{figure*}

In \autoref{fig:symmtery} we show how different twisting of tetrahedral chains leads to polar, nonpolar or antipolar structures
 \cite{Luo/Hayward2013,Young2015,Tian2018,Parsons2009}. Our total energy calculations for  \NCFO and \CFO
show that polar structures can only be stabilized for Grenier phases, while antipolar structures can only be stabilized in browmillerite phases (see \autoref{table:energetics}).
These results point at the key role of oxygen vacancy concentration ($\delta$) in oxygen-deficient perovskites ($AB$O$_{3-\delta}$) to obtain the desired  ferroelectric behavior, specifically, a  $\delta$ value of $\nicefrac{1}{3}$ is required to realize polar phases.

Interestingly, we find that cooperative distortions are responsible for the stabilization of  a polar (antipolar) phase in Grenier (brownmillerite) structures.
In particular, as shown in \autoref{fig:distortion},
 rotations (bending) of octahedral units lead to antipolar  (polar) structures in browmillerite, while the opposite happens in Grenier phases, where rotations (bending) of the  units lead to polar (antipolar) structures.

One may intuitively understand the emergence of ferroelectricity in Grenier  phases by  drawing an analogy between cooperative distortions and gear motion, as shown in
\autoref{fig:gear}.
Upon rotation of octahedral  and twisting of tetrahedral units,
the apical oxygen atoms of the octahedral units move in  opposite directions, whereas
the ones in the tetrahedral units move in the same direction [\autoref{fig:gear}(a,b)].
We may then view octahedral layers as rotating gears, and  tetrahedral layers as sliding gears. Hence ferroelectricity arises from the combination of the rotational and sliding motion of octahedra and tetrahedra, respectively.
We note that these cooperative distortions occur without any off-center displacement of Fe atoms, consistent with the  general trend observed for systems with partially filled $d$-orbitals
\cite{Spaldin2000}.

%

\begin{figure}[h]
\centering
\includegraphics[width=0.85\columnwidth]{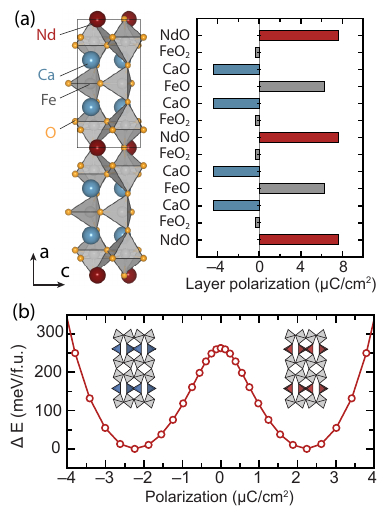}
\caption{
\textbf{Polarization and switching barrier in \NCFO.}
(a) Layer-by-layer projection of the polarization in the \NCFO Grenier phase (see text). 
(b) Total energy difference between two polar phases with opposite polarizations as a function of the polarization.
}
\label{fig:polarization}
\end{figure}

Having identified which structural motifs lead to polar phases, we calculated the spontaneous polarization of the solids using  DFT+$U$ and the Berry-phase method \cite{Spaldin2012,Resta/Vanderbilt2007}.
For \NCFO we obtained 2.24 $\mu\mathrm{C/cm}^2$, a value which is insensitive to the choice of the $U$ value, with variations within 0.11 \Ccm when $U$ is varied between 4 to 7 eV. (SI). This value is smaller than that reported experimentally \cite{Shimakawa/Martin2019} for $R_{1.2}$Ba$_{1.2}$Ca$_{0.6}$Fe$_3$O$_8$ ($R$ = Tb, Gd), most likely due to the difference of $R$ and $A$ cations, as discussed in the next section.
In \autoref{fig:polarization}(b), we show the difference in polarization  between the nonpolar and the polar phases. Assuming that the path identified here is indeed the lowest energy path, we find that the energy barrier to switch the sign of the polarization is 263 meV/f.u. for \NCFO.  
This barrier is larger than that found in \SFO $\sim$150 meV/f.u. \cite{Kang/Choi2019}), possibly because of smaller $R$-ions and $A$-cations.

In order to better understand the mechanism leading to ferroelectricity, it is useful to analyze the contributions of each layer to the total polarization.
To this end, 
we plot the layer polarization along the polar ($z$) direction in \autoref{fig:polarization}(a),  computed by summing, for each constituent ion ($n$), the products of Born effective charges ($Z^*$) and the displacement of ions from the centro-symmetric site ($\Delta u$) along the polar direction:
\begin{equation}
P_z = \frac{e}{\Omega}\sum_n Z^*_{zz,n}\Delta u_{z,n},
\label{eq:p}
\end{equation}
Here $e$ is the electron charge and $\Omega$ is the volume of the unit cell. As shown in \autoref{fig:polarization}(a),
the tetrahedral chains give rise to local dipole moments, with additional, significant contributions of opposite sign from CaO layers (negative contribution) and NdO layers (positive contribution).

We note that our results are  qualitatively different from those of the model suggested in Ref.~\onlinecite{Shimakawa/Martin2019},
which predicts positive contributions to the polarization from all atomic layers and contributions from $R$O and $A$O layers smaller than those found here. In the model of Ref.~\onlinecite{Shimakawa/Martin2019} all apical oxygen atoms are displaced in the same direction.

The  ferroelectricity found here in \NCFO Grenier phases originates from the presence of {\it two} octahedral layers between pairs of tetrahedral layers, i.e. the emergence of ferroelectricity is critically dependent not only on oxygen deficiency but also on the stacking of layers, which differs from the stacking found in brownmillerite composed only of {\it one} octahedral layer between pairs of tetrahedral ones. Our results also indicate that 
 any even number of intercalated octahedral layers would lead to the stabilization of a polar phase.
However, to the best of our knowledge, there is no compound with a number of intercalated octahedral layers larger than two (as in Grenier structures); hence we consider Grenier structures as the optimal solution for designing ferroelectricity in the class of materials studied here, and in the next section we consider only solids with the Grenier stacking.%


\subsection{Cation size-effect in ferroelectricity}
\label{ssec:cation}

We now turn to discuss  the effect of the size of  $R$-ions and $A$-cations in determining the ferroelectric properties of \RAFO. 
In \autoref{fig:symmtery}  we consider a cation ordering in which the $R$-ions are located between the octahedral layers because our total energy calculations as a function of 
cation orderings in \NCFO showed that the rare earth preferably occupies the $A$-site in octahedral units (see SI).
The ordering chosen here is also consistent with the cation occupancies reported in the literature for  \LSFO and \NCFO \cite{Hudspeth2009}.

\begin{figure}[h]
\centering
\includegraphics[width=0.85\columnwidth]{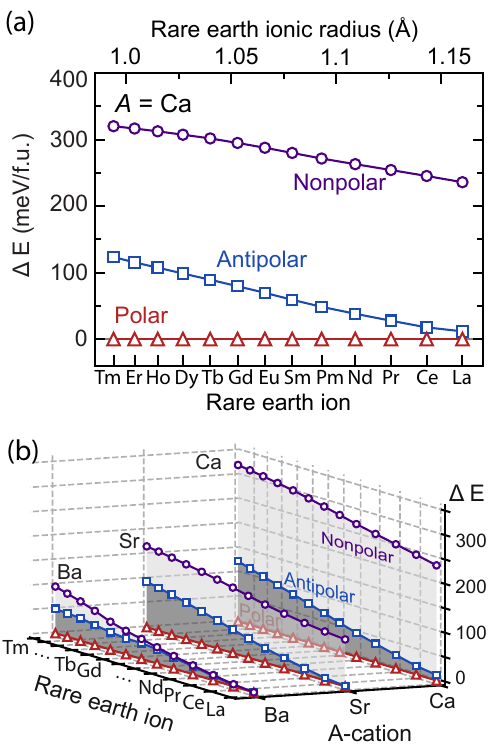}
\caption{
\textbf{Cation-dependent energy differences between distortion patterns in Grenier structure.}
(a) Energy difference between nonpolar, antipolar and polar phases in Grenier structures of $R_{1/3}$Ca$_{2/3}$FeO$_{2.67}$ as a function of $R$-ions. 
(b) Total energy difference between nonpolar, antipolar, and polar phases as a function of $A$-cation and $R$-ions.
Two-dimensional projections for $A$ = Sr and Br are provided in SI.
}
\label{fig:cation_effect}
\end{figure}

In \autoref{fig:cation_effect}, we show the total energy difference between polar and nonpolar phases as a function of $R$ and $A$.
For all \RAFO systems we find that the polar phase is more stable than the antipolar one
and  small-size $R$ and $A$ contribute to the stabilization of the polar phase.
The systems with the smallest cations ($R$ = Tm, $A$ = Ca) exhibit the largest energy differences between different phases,
and the polar phase is 123 and 320 meV/f.u. lower than antipolar and nonpolar phases, respectively.
On the other hand, for larger cations (e.g. $R$ = La, $A$ = Ba) we find that these energy differences almost vanish (they are less than 1 meV/f.u.), indicating that in the Grenier structure all phases  are nearly degenerate.

%


Interestingly,  we find that for all $R$-ions, only the polar phase can be stabilized.
In \autoref{fig:cation_effect}, we show that the energy difference between antipolar  and nonpolar phases is not sensitive to the size of the $R$-ion, since the local geometries surrounding the element $R$ are similar in the nonpolar and antipolar phases, whose main difference is  the twisting of the tetrahedral chains. 
For example in systems with $A$ = Ca  shown in \autoref{fig:cation_effect}(a), the energy difference between nonpolar and antipolar phases shows a weak variation from 197 meV/f.u. for $R$ = Tm  to 224 meV/f.u. for $R$ = La.
On the other hand, the energy differences between nonpolar, polar, and antipolar phases vary substantially as a function of  the $A$ cation. The energy difference decreases by $50 \%$ when going from from Ca to Sr, and is again halved in the case of Ba.

\begin{figure}[ht]
\centering
\includegraphics[width=0.90\columnwidth]{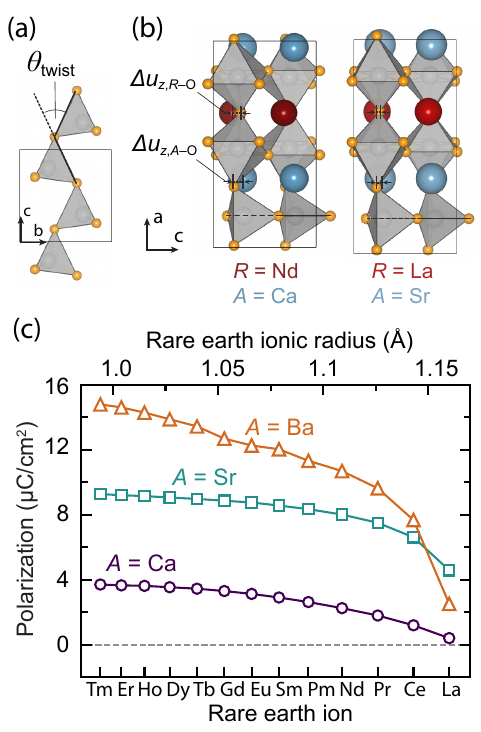}
\caption{
\textbf{Geometric descriptors of the polar \RAFO phases and the effect of cation size in polarization.}
(a) Twist angle ($\theta_\textrm{twist}$) of tetrahedral chains  in polar Grenier phase of \NCFO. 
(b) Atomic structure and polar displacements in $R$O and $A$O layers in polar Grenier phases 
of \NCFO and \LSFO.
(b) Spontaneous polarization of \RAFO as a function of the rare-earth ion.
}
\label{fig:P_cation}
\end{figure}

Based on our results on total  energy differences and the layer-by-layer projection of the polarization presented in \autoref{fig:polarization}(a),
we identify three major geometrical descriptors for the emergence of the polarization in \autoref{fig:P_cation}(a,b): the twist angle of tetrahedral chains ($\theta_\textrm{twist}$) and the displacements of $R-\rm{O}$ pairs along the polar direction ($\Delta u_{z,R-\rm{O}}$) and of $A-\rm{O}$ pairs ($\Delta u_{z,A-\rm{O}}$). 
We found that these geometrical parameters undergo substantial variations,  depending on the choice of $R$ and $A$ (see SI).
For example, for 
\NCFO (small cation), $\Delta u_{z,R-\rm{O}}$ =  0.45 \AA\xspace and $\Delta u_{z,A-\rm{O}}$ = 0.47 \AA, while for \LSFO (large cation), those values are smaller,  0.14 and 0.19 \AA, respectively, as shown in \autoref{fig:P_cation}(b).
On the other hand, the difference in $\theta_\textrm{twist}$ is relatively small, with the angle of \NCFO being 51.5\degree and that of \LSFO  48.7\degree.
%

%
In \autoref{fig:P_cation}(b), we show the polarization as a function of $R$ and $A$, exhibiting a  maximum for 
$R$ = Tm and $A$ = Ba.
This cation selection gives rise to the largest $\Delta u_{z,R-\rm{O}}$ and smallest $\Delta u_{z,A-\rm{O}}$, which in turn maximize the spontaneous polarization.
We note that the value found here for the polarization is  lower than the values reported experimentally, ranging from 23.2 to 33.0 \Ccm \cite{Shimakawa/Martin2019}.
Different reasons may be responsible for this discrepancy. From a theoretical standpoint,  the absolute value of $P$ is expected to depend on the choice of the functionals \cite{Yuk2017}, and the trend shown in \autoref{fig:P_cation}, more than the absolute values, should be considered to define a reliable roadmap for ferroelectric design.
It is noteworthy though that the polarization values computed here are not sensitive to the choice of the $U$ value (see SI).
We also note that the chemical formula of $R_{1.2}$Ba$_{1.2}$Ca$_{0.6}$Fe$_3$O$_8$ ($R$ = Gd, Tb) reported experimentally deviates from that of the \textit{stoichiometric} \RAFO  considered here,  and may involve cation disorder.
We conclude that smaller $R$ such as Tm and larger $A$ such as Ba are optimal choices for enhancing ferroelectricity of  polar Grenier structures.
In addition, given that the primary switching barrier is formed between the polar and nonpolar phases, we find that small $A$-cations, e.g. Ca, lead to a high energy barrier.
While our results show that the energy barrier also changes depending on the choice of $R$,
the polarization value does not substantially vary when $R$ is smaller than Gd [\autoref{fig:P_cation}(b)]. Specifically, changing $R$ from the smallest member (Tm) of the set to the slightly larger Tb in $R_{1/3}$Ba$_{2/3}$FeO$_{2.67}$ would decrease the spontaneous polarization from 14.8 to 13.4 \Ccm. For $A$ = Sr and Ca, the decrease in polarization is 0.30 and 0.26 \Ccm, respectively. This small variation of $P$ indicates that several $R$ elements such as Tb, Dy, and Ho may be chosen to obtain a similar ferroelectric behaviour in Grenier structures.

\section {Discussion}

Using the results of first-principles calculations we proposed that Grenier structures arising from oxygen-vacancy ordering in 3D oxide perovskites are interesting materials to realize ferroelectricity.
In Grenier structures polar phases are stabilized by the cooperative motion of octahedral and tetrahedral units, with the twisting motion of tetrahedral chains playing a key role in forming local dipole moments.
We considered in particular $R_{1/3}A_{2/3}$FeO$_{2.67}$ solids, where the oxidation state of Fe is +3,  and whose composition is optimal to grow Grenier rather than brownmillerite structures \cite{Hautier2017}. We investigated the emergence of ferroelectricity in \RAFO as a function  of the size of the rare-earth and alkali-earth elements, and provided a strategy to choose optimal elements to obtain polar phases.
Interestingly, Grenier structures can also be present  in superlattices.
For example in Ref.~\onlinecite{Mishra2014}, the introduction of oxygen vacancies in (LaFeO$_3$)$_2$/(SrFeO$_3$) superlattices was shown to form polar domains. Despite a different ratio of La to Sr compared to the ratios considered here, it is reasonable to expect that Grenier phases formed at interfaces may be responsible for the polar domains observed in the films.

We emphasize that the mechanism leading to ferroelectricity in Grenier phases, which exhibits three-dimensional connectivities of polyhedra, is distinct from that responsible for the same phenomenon in other perovskite materials.
For example, in  the layered Ruddlesden-Popper and Dion-Jacobson phases \cite{HIF:Benedek2011,DJ:Benedek2014}, octahedra are connected in only two dimensions, and a hybrid improper mechanism is responsible \cite{HIF:Benedek2011,Lu/Rondinelli2016,Benedek2015} for ferroelectricity, where two nonpolar rotational modes of octahedral units give rise to a finite polarization. Other materials with Ruddlesden-Popper phases, for example hybrid organic-inorganic halide perovskites \cite{Zhang/Chien2019} exhibit ferroelectricity  arising from the anisotropic shape of organic molecules, which leads to distortions of the octahedral units. The ferroelectricity of Grenier phases also differs from that of oxygen-deficient oxides with defect-like vacancies, e.g. in SrMnO$_{3-\delta}$ and SmFeO$_{3-\delta}$ \cite{Becher/Manfred2015,Li/Chen2022}. 

Overall, our work suggests that Grenier phases can be a rich platform for novel materials phenomena arising from breaking inversion symmetry in solids.
The geometrically driven ferroelectric behavior observed in Grenier phases can be extended to other $d$-electron configurations by adjusting the $R$/$A$ ratio or incorporating different transition metals. Notably, oxides containing elements such as Mn, Al, Ga, and In, which are known to be compatible with tetrahedral coordination, can form Grenier or brownmillerite structures \cite{Karki2020,Kahlenberg2000,Marik/LeBreton2019,Sr2In2O5}.
By exploring the chemistry of polar Grenier phases, we anticipate exciting opportunities to realize materials with  multiferroic properties and investigate optical coupling with ferroelectricity. This line of inquiry could pave the way for utilizing Grenier phases in energy-efficient computing devices, where the inherent hysteresis of ferroelectricity can help realize nonvolatile memories for neuromorphic devices \cite{Spaldin_review2019,Khan2020}.

\section{Methods}

We performed density function theory (DFT) calculations with the generalized gradient corrected functional proposed by Perdew-Burke-Ernzerhof (PBE) \cite{PBE}, and using the plane wave code Quantum ESPRESSO (QE) \cite{Giannozzi_2009,Giannozzi_2017,Giannozzi_2020}, with a $U$ parameter \cite{Cococcioni2005} for Fe of 6.5 eV. This $U$ value was determined by matching the experimental band gap of the  SrFeO$_{2.5}$ brownmillerites \cite{Galakhov/Kozhevnikov2010}.
We find the calculated total energies and polarization values are relatively insensitive  to the choice of $U$ (see SI).
We used projected augmented wave (PAW) pseudopotentials from the PSlibrary \cite{DalCorso2014},
except for La where we used the Rare Earth PAW datasets \cite{Topsakal2014}.
For rare-earth elements, $f$-electrons were considered as part of the core and we used pseudopotentials  having 11 valence electrons.
assuming a 3+ charge state of each element.
We used a kinetic energy cutoff of 75 Ry for wavefunctions and 300 Ry for the charge density.
The Brillouin zone was sampled with a $3\times 8 \times 8$ $k$-grid for Grenier structures whose unit cell size is $6a_{\rm pc} \times 2\sqrt{2}a_{\rm pc} \times 2\sqrt{2}a_{\rm pc}$, where $a_{\rm pc}$ denotes the pseudocubic lattice parameter of perovskites.
The cell parameters and atomic coordinates were optimized with a threshold of 3$\times$10$^{-4}$ Ry/Bohr ($\sim$ 0.01 eV/\AA) on the atomic forces.
We carried out calculations for  \RAFO solids with different choices of $R$ and $A$ elements:$R$ is a lanthanide elements from La to Tm, where $R$FeO$_3$ solids are known to form perovskite structures \cite{Zhang/Ansermet2021,Aparnadevi2016,Slawinski2005,Ritter2022,Ritter2021,Prakash2019,Cao2014,Bertaut1967};
 $A$ = Ca, Sr, and Ba.
%
When comparing the size of rare-earth cations, we used 8-coordinate ionic radii \cite{Shannon1976},
similar to nickelate perovskite studies \cite{Wagner/Rondinelli2018,Guo/Khomskii2018}.
Although the $R$ cations are expected to be 12-fold coordinated in cubic perovskites, the
nickelates possess GdFeO$_3$-type distortions, which reduce the $R$ cation coordination. 
Visualization of atomic structures was done with the VESTA package \cite{VESTA}.


\begin{acknowledgments}
The authors would like to acknowledge valuable discussions with James M. Rondinelli, Danilo Puggioni, and Shenli Zhang. 
This research was conducted as part of the Quantum Materials for Energy Efficient Neuromorphic Computing, an Energy Frontier Research Center funded by the US Department of Energy, Office of Science, Basic Energy Sciences under award DE-SC0019273.
This research used computational resources of the University of Chicago's Research Computing Center and at the National Energy Research Scientific Computing Center (NERSC), a DOE Office of Science User Facility supported by the Office of Science of the US Department of Energy.
\end{acknowledgments}

\section*{Data availability}
Data that support the findings of this study will be available through the Qresp \cite{Qresp} curator.

\section*{Author contributions}
Both authors designed the research and wrote the manuscript. Y.S.performed all calculations.

\section*{Competing interests}
The authors declare no competing financial or non-financial interests.


\begin{thebibliography}{80}%
\makeatletter
\providecommand \@ifxundefined [1]{%
 \@ifx{#1\undefined}
}%
\providecommand \@ifnum [1]{%
 \ifnum #1\expandafter \@firstoftwo
 \else \expandafter \@secondoftwo
 \fi
}%
\providecommand \@ifx [1]{%
 \ifx #1\expandafter \@firstoftwo
 \else \expandafter \@secondoftwo
 \fi
}%
\providecommand \natexlab [1]{#1}%
\providecommand \enquote  [1]{``#1''}%
\providecommand \bibnamefont  [1]{#1}%
\providecommand \bibfnamefont [1]{#1}%
\providecommand \citenamefont [1]{#1}%
\providecommand \href@noop [0]{\@secondoftwo}%
\providecommand \href [0]{\begingroup \@sanitize@url \@href}%
\providecommand \@href[1]{\@@startlink{#1}\@@href}%
\providecommand \@@href[1]{\endgroup#1\@@endlink}%
\providecommand \@sanitize@url [0]{\catcode `\\12\catcode `\$12\catcode
  `\&12\catcode `\#12\catcode `\^12\catcode `\_12\catcode `\%12\relax}%
\providecommand \@@startlink[1]{}%
\providecommand \@@endlink[0]{}%
\providecommand \url  [0]{\begingroup\@sanitize@url \@url }%
\providecommand \@url [1]{\endgroup\@href {#1}{\urlprefix }}%
\providecommand \urlprefix  [0]{URL }%
\providecommand \Eprint [0]{\href }%
\providecommand \doibase [0]{https://doi.org/}%
\providecommand \selectlanguage [0]{\@gobble}%
\providecommand \bibinfo  [0]{\@secondoftwo}%
\providecommand \bibfield  [0]{\@secondoftwo}%
\providecommand \translation [1]{[#1]}%
\providecommand \BibitemOpen [0]{}%
\providecommand \bibitemStop [0]{}%
\providecommand \bibitemNoStop [0]{.\EOS\space}%
\providecommand \EOS [0]{\spacefactor3000\relax}%
\providecommand \BibitemShut  [1]{\csname bibitem#1\endcsname}%
\let\auto@bib@innerbib\@empty
\bibitem [{\citenamefont {Ramesh}\ and\ \citenamefont
  {Schlom}(2019)}]{Ramesh/Schlom2019}%
  \BibitemOpen
  \bibfield  {author} {\bibinfo {author} {\bibfnamefont {R.}~\bibnamefont
  {Ramesh}}\ and\ \bibinfo {author} {\bibfnamefont {D.~G.}\ \bibnamefont
  {Schlom}},\ }\bibfield  {title} {\bibinfo {title} {Creating emergent
  phenomena in oxide superlattices},\ }\href
  {https://doi.org/10.1038/s41578-019-0095-2} {\bibfield  {journal} {\bibinfo
  {journal} {Nature Reviews Materials}\ }\textbf {\bibinfo {volume} {4}},\
  \bibinfo {pages} {257} (\bibinfo {year} {2019})}\BibitemShut {NoStop}%
\bibitem [{\citenamefont {Sicron}\ \emph {et~al.}(1994)\citenamefont {Sicron},
  \citenamefont {Ravel}, \citenamefont {Yacoby}, \citenamefont {Stern},
  \citenamefont {Dogan},\ and\ \citenamefont {Rehr}}]{Sicron1994}%
  \BibitemOpen
  \bibfield  {author} {\bibinfo {author} {\bibfnamefont {N.}~\bibnamefont
  {Sicron}}, \bibinfo {author} {\bibfnamefont {B.}~\bibnamefont {Ravel}},
  \bibinfo {author} {\bibfnamefont {Y.}~\bibnamefont {Yacoby}}, \bibinfo
  {author} {\bibfnamefont {E.~A.}\ \bibnamefont {Stern}}, \bibinfo {author}
  {\bibfnamefont {F.}~\bibnamefont {Dogan}},\ and\ \bibinfo {author}
  {\bibfnamefont {J.~J.}\ \bibnamefont {Rehr}},\ }\bibfield  {title} {\bibinfo
  {title} {Nature of the ferroelectric phase transition in {PbTiO}$_{3}$},\
  }\href {https://doi.org/10.1103/PhysRevB.50.13168} {\bibfield  {journal}
  {\bibinfo  {journal} {Phys. Rev. B}\ }\textbf {\bibinfo {volume} {50}},\
  \bibinfo {pages} {13168} (\bibinfo {year} {1994})}\BibitemShut {NoStop}%
\bibitem [{\citenamefont {Stern}(2004)}]{Stern2004}%
  \BibitemOpen
  \bibfield  {author} {\bibinfo {author} {\bibfnamefont {E.~A.}\ \bibnamefont
  {Stern}},\ }\bibfield  {title} {\bibinfo {title} {Character of order-disorder
  and displacive components in barium titanate},\ }\href
  {https://doi.org/10.1103/PhysRevLett.93.037601} {\bibfield  {journal}
  {\bibinfo  {journal} {Phys. Rev. Lett.}\ }\textbf {\bibinfo {volume} {93}},\
  \bibinfo {pages} {037601} (\bibinfo {year} {2004})}\BibitemShut {NoStop}%
\bibitem [{\citenamefont {Khan}\ \emph {et~al.}(2014)\citenamefont {Khan},
  \citenamefont {Chatterjee}, \citenamefont {Wang}, \citenamefont {Drapcho},
  \citenamefont {You}, \citenamefont {Serrao}, \citenamefont {Bakaul},
  \citenamefont {Ramesh},\ and\ \citenamefont {Salahuddin}}]{Khan2014}%
  \BibitemOpen
  \bibfield  {author} {\bibinfo {author} {\bibfnamefont {A.~I.}\ \bibnamefont
  {Khan}}, \bibinfo {author} {\bibfnamefont {K.}~\bibnamefont {Chatterjee}},
  \bibinfo {author} {\bibfnamefont {B.}~\bibnamefont {Wang}}, \bibinfo {author}
  {\bibfnamefont {S.}~\bibnamefont {Drapcho}}, \bibinfo {author} {\bibfnamefont
  {L.}~\bibnamefont {You}}, \bibinfo {author} {\bibfnamefont {C.}~\bibnamefont
  {Serrao}}, \bibinfo {author} {\bibfnamefont {S.~R.}\ \bibnamefont {Bakaul}},
  \bibinfo {author} {\bibfnamefont {R.}~\bibnamefont {Ramesh}},\ and\ \bibinfo
  {author} {\bibfnamefont {S.}~\bibnamefont {Salahuddin}},\ }\bibfield  {title}
  {\bibinfo {title} {Negative capacitance in a ferroelectric capacitor},\
  }\href {https://doi.org/10.1038/nmat4148} {\bibfield  {journal} {\bibinfo
  {journal} {Nature Materials}\ }\textbf {\bibinfo {volume} {14}},\ \bibinfo
  {pages} {182} (\bibinfo {year} {2014})}\BibitemShut {NoStop}%
\bibitem [{\citenamefont {Lines}\ and\ \citenamefont
  {Glass}(1977)}]{Lines1977}%
  \BibitemOpen
  \bibfield  {author} {\bibinfo {author} {\bibfnamefont {M.~E.}\ \bibnamefont
  {Lines}}\ and\ \bibinfo {author} {\bibfnamefont {A.~M.}\ \bibnamefont
  {Glass}},\ }\href@noop {} {\emph {\bibinfo {title} {Principles and
  applications of ferroelectrics and related materials}}}\ (\bibinfo
  {publisher} {Oxford university press},\ \bibinfo {year} {1977})\BibitemShut
  {NoStop}%
\bibitem [{\citenamefont {{\'{I}}{\~{n}}iguez}\ \emph
  {et~al.}(2019)\citenamefont {{\'{I}}{\~{n}}iguez}, \citenamefont {Zubko},
  \citenamefont {Luk'yanchuk},\ and\ \citenamefont {Cano}}]{Iniguez2019}%
  \BibitemOpen
  \bibfield  {author} {\bibinfo {author} {\bibfnamefont {J.}~\bibnamefont
  {{\'{I}}{\~{n}}iguez}}, \bibinfo {author} {\bibfnamefont {P.}~\bibnamefont
  {Zubko}}, \bibinfo {author} {\bibfnamefont {I.}~\bibnamefont {Luk'yanchuk}},\
  and\ \bibinfo {author} {\bibfnamefont {A.}~\bibnamefont {Cano}},\ }\bibfield
  {title} {\bibinfo {title} {Ferroelectric negative capacitance},\ }\href
  {https://doi.org/10.1038/s41578-019-0089-0} {\bibfield  {journal} {\bibinfo
  {journal} {Nature Reviews Materials}\ }\textbf {\bibinfo {volume} {4}},\
  \bibinfo {pages} {243} (\bibinfo {year} {2019})}\BibitemShut {NoStop}%
\bibitem [{\citenamefont {Bersuker}(2013)}]{Bersuker2013}%
  \BibitemOpen
  \bibfield  {author} {\bibinfo {author} {\bibfnamefont {I.~B.}\ \bibnamefont
  {Bersuker}},\ }\bibfield  {title} {\bibinfo {title} {{Pseudo-Jahn–Teller}
  effect—a two-state paradigm in formation, deformation, and transformation
  of molecular systems and solids},\ }\href {https://doi.org/10.1021/cr300279n}
  {\bibfield  {journal} {\bibinfo  {journal} {Chemical Reviews}\ }\textbf
  {\bibinfo {volume} {113}},\ \bibinfo {pages} {1351} (\bibinfo {year}
  {2013})}\BibitemShut {NoStop}%
\bibitem [{\citenamefont {Hill}(2000)}]{Spaldin2000}%
  \BibitemOpen
  \bibfield  {author} {\bibinfo {author} {\bibfnamefont {N.~A.}\ \bibnamefont
  {Hill}},\ }\bibfield  {title} {\bibinfo {title} {Why are there so few
  magnetic ferroelectrics?},\ }\href {https://doi.org/10.1021/jp000114x}
  {\bibfield  {journal} {\bibinfo  {journal} {The Journal of Physical Chemistry
  B}\ }\textbf {\bibinfo {volume} {104}},\ \bibinfo {pages} {6694} (\bibinfo
  {year} {2000})}\BibitemShut {NoStop}%
\bibitem [{\citenamefont {Spaldin}\ and\ \citenamefont
  {Ramesh}(2019)}]{Spaldin_review2019}%
  \BibitemOpen
  \bibfield  {author} {\bibinfo {author} {\bibfnamefont {N.~A.}\ \bibnamefont
  {Spaldin}}\ and\ \bibinfo {author} {\bibfnamefont {R.}~\bibnamefont
  {Ramesh}},\ }\bibfield  {title} {\bibinfo {title} {Advances in
  magnetoelectric multiferroics},\ }\href
  {https://doi.org/10.1038/s41563-018-0275-2} {\bibfield  {journal} {\bibinfo
  {journal} {Nature Materials}\ }\textbf {\bibinfo {volume} {18}},\ \bibinfo
  {pages} {203} (\bibinfo {year} {2019})}\BibitemShut {NoStop}%
\bibitem [{\citenamefont {Bousquet}\ \emph {et~al.}(2008)\citenamefont
  {Bousquet}, \citenamefont {Dawber}, \citenamefont {Stucki}, \citenamefont
  {Lichtensteiger}, \citenamefont {Hermet}, \citenamefont {Gariglio},
  \citenamefont {Triscone},\ and\ \citenamefont {Ghosez}}]{Bousquet2008}%
  \BibitemOpen
  \bibfield  {author} {\bibinfo {author} {\bibfnamefont {E.}~\bibnamefont
  {Bousquet}}, \bibinfo {author} {\bibfnamefont {M.}~\bibnamefont {Dawber}},
  \bibinfo {author} {\bibfnamefont {N.}~\bibnamefont {Stucki}}, \bibinfo
  {author} {\bibfnamefont {C.}~\bibnamefont {Lichtensteiger}}, \bibinfo
  {author} {\bibfnamefont {P.}~\bibnamefont {Hermet}}, \bibinfo {author}
  {\bibfnamefont {S.}~\bibnamefont {Gariglio}}, \bibinfo {author}
  {\bibfnamefont {J.-M.}\ \bibnamefont {Triscone}},\ and\ \bibinfo {author}
  {\bibfnamefont {P.}~\bibnamefont {Ghosez}},\ }\bibfield  {title} {\bibinfo
  {title} {Improper ferroelectricity in perovskite oxide artificial
  superlattices},\ }\href {https://doi.org/10.1038/nature06817} {\bibfield
  {journal} {\bibinfo  {journal} {Nature}\ }\textbf {\bibinfo {volume} {452}},\
  \bibinfo {pages} {732} (\bibinfo {year} {2008})}\BibitemShut {NoStop}%
\bibitem [{\citenamefont {Fennie}\ and\ \citenamefont
  {Rabe}(2005)}]{Fennie/Rabe2005}%
  \BibitemOpen
  \bibfield  {author} {\bibinfo {author} {\bibfnamefont {C.~J.}\ \bibnamefont
  {Fennie}}\ and\ \bibinfo {author} {\bibfnamefont {K.~M.}\ \bibnamefont
  {Rabe}},\ }\bibfield  {title} {\bibinfo {title} {Ferroelectric transition in
  {YMnO}$_3$ from first principles},\ }\href
  {https://doi.org/10.1103/PhysRevB.72.100103} {\bibfield  {journal} {\bibinfo
  {journal} {Phys. Rev. B}\ }\textbf {\bibinfo {volume} {72}},\ \bibinfo
  {pages} {100103} (\bibinfo {year} {2005})}\BibitemShut {NoStop}%
\bibitem [{\citenamefont {Benedek}\ and\ \citenamefont
  {Fennie}(2011)}]{HIF:Benedek2011}%
  \BibitemOpen
  \bibfield  {author} {\bibinfo {author} {\bibfnamefont {N.~A.}\ \bibnamefont
  {Benedek}}\ and\ \bibinfo {author} {\bibfnamefont {C.~J.}\ \bibnamefont
  {Fennie}},\ }\bibfield  {title} {\bibinfo {title} {Hybrid improper
  ferroelectricity: A mechanism for controllable polarization-magnetization
  coupling},\ }\href {https://doi.org/10.1103/PhysRevLett.106.107204}
  {\bibfield  {journal} {\bibinfo  {journal} {Phys. Rev. Lett.}\ }\textbf
  {\bibinfo {volume} {106}},\ \bibinfo {pages} {107204} (\bibinfo {year}
  {2011})}\BibitemShut {NoStop}%
\bibitem [{\citenamefont {Benedek}\ \emph {et~al.}(2015)\citenamefont
  {Benedek}, \citenamefont {Rondinelli}, \citenamefont {Djani}, \citenamefont
  {Ghosez},\ and\ \citenamefont {Lightfoot}}]{Benedek2015}%
  \BibitemOpen
  \bibfield  {author} {\bibinfo {author} {\bibfnamefont {N.~A.}\ \bibnamefont
  {Benedek}}, \bibinfo {author} {\bibfnamefont {J.~M.}\ \bibnamefont
  {Rondinelli}}, \bibinfo {author} {\bibfnamefont {H.}~\bibnamefont {Djani}},
  \bibinfo {author} {\bibfnamefont {P.}~\bibnamefont {Ghosez}},\ and\ \bibinfo
  {author} {\bibfnamefont {P.}~\bibnamefont {Lightfoot}},\ }\bibfield  {title}
  {\bibinfo {title} {Understanding ferroelectricity in layered perovskites: new
  ideas and insights from theory and experiments},\ }\href
  {https://doi.org/10.1039/C5DT00010F} {\bibfield  {journal} {\bibinfo
  {journal} {Dalton Transactions}\ }\textbf {\bibinfo {volume} {44}},\ \bibinfo
  {pages} {10543} (\bibinfo {year} {2015})}\BibitemShut {NoStop}%
\bibitem [{\citenamefont {Rondinelli}\ and\ \citenamefont
  {May}(2019)}]{Rondinelli/May2019}%
  \BibitemOpen
  \bibfield  {author} {\bibinfo {author} {\bibfnamefont {J.~M.}\ \bibnamefont
  {Rondinelli}}\ and\ \bibinfo {author} {\bibfnamefont {S.~J.}\ \bibnamefont
  {May}},\ }\bibfield  {title} {\bibinfo {title} {Deliberate deficiencies:
  Expanding electronic function through non-stoichiometry},\ }\href
  {https://doi.org/https://doi.org/10.1016/j.matt.2019.06.013} {\bibfield
  {journal} {\bibinfo  {journal} {Matter}\ }\textbf {\bibinfo {volume} {1}},\
  \bibinfo {pages} {33} (\bibinfo {year} {2019})}\BibitemShut {NoStop}%
\bibitem [{\citenamefont {Anderson}\ \emph {et~al.}(1993)\citenamefont
  {Anderson}, \citenamefont {Vaughey},\ and\ \citenamefont
  {Poeppelmeier}}]{Anderson1993}%
  \BibitemOpen
  \bibfield  {author} {\bibinfo {author} {\bibfnamefont {M.~T.}\ \bibnamefont
  {Anderson}}, \bibinfo {author} {\bibfnamefont {J.~T.}\ \bibnamefont
  {Vaughey}},\ and\ \bibinfo {author} {\bibfnamefont {K.~R.}\ \bibnamefont
  {Poeppelmeier}},\ }\bibfield  {title} {\bibinfo {title} {Structural
  similarities among oxygen-deficient perovskites},\ }\href
  {https://doi.org/10.1021/cm00026a003} {\bibfield  {journal} {\bibinfo
  {journal} {Chemistry of Materials}\ }\textbf {\bibinfo {volume} {5}},\
  \bibinfo {pages} {151} (\bibinfo {year} {1993})}\BibitemShut {NoStop}%
\bibitem [{\citenamefont {Chaturvedi}\ \emph {et~al.}(2021)\citenamefont
  {Chaturvedi}, \citenamefont {Postiglione}, \citenamefont {Chakraborty},
  \citenamefont {Yu}, \citenamefont {Tabiś}, \citenamefont {Hameed},
  \citenamefont {Biniskos}, \citenamefont {Jacobson}, \citenamefont {Zhang},
  \citenamefont {Zhou}, \citenamefont {Greven}, \citenamefont {Ferry},\ and\
  \citenamefont {Leighton}}]{Chaturvedi/Leighton2021}%
  \BibitemOpen
  \bibfield  {author} {\bibinfo {author} {\bibfnamefont {V.}~\bibnamefont
  {Chaturvedi}}, \bibinfo {author} {\bibfnamefont {W.~M.}\ \bibnamefont
  {Postiglione}}, \bibinfo {author} {\bibfnamefont {R.~D.}\ \bibnamefont
  {Chakraborty}}, \bibinfo {author} {\bibfnamefont {B.}~\bibnamefont {Yu}},
  \bibinfo {author} {\bibfnamefont {W.}~\bibnamefont {Tabiś}}, \bibinfo
  {author} {\bibfnamefont {S.}~\bibnamefont {Hameed}}, \bibinfo {author}
  {\bibfnamefont {N.}~\bibnamefont {Biniskos}}, \bibinfo {author}
  {\bibfnamefont {A.}~\bibnamefont {Jacobson}}, \bibinfo {author}
  {\bibfnamefont {Z.}~\bibnamefont {Zhang}}, \bibinfo {author} {\bibfnamefont
  {H.}~\bibnamefont {Zhou}}, \bibinfo {author} {\bibfnamefont {M.}~\bibnamefont
  {Greven}}, \bibinfo {author} {\bibfnamefont {V.~E.}\ \bibnamefont {Ferry}},\
  and\ \bibinfo {author} {\bibfnamefont {C.}~\bibnamefont {Leighton}},\
  }\bibfield  {title} {\bibinfo {title} {{Doping- and Strain-Dependent
  Electrolyte-Gate-Induced Perovskite to Brownmillerite Transformation in
  Epitaxial La$_{1–x}$Sr$_x$CoO$_{3−\delta}$ Films}},\ }\href
  {https://doi.org/10.1021/acsami.1c13828} {\bibfield  {journal} {\bibinfo
  {journal} {ACS Applied Materials \& Interfaces}\ }\textbf {\bibinfo {volume}
  {13}},\ \bibinfo {pages} {51205} (\bibinfo {year} {2021})}\BibitemShut
  {NoStop}%
\bibitem [{\citenamefont {Walter}\ \emph {et~al.}(2020)\citenamefont {Walter},
  \citenamefont {Bose}, \citenamefont {Cabero}, \citenamefont {Varela},\ and\
  \citenamefont {Leighton}}]{Leighton2020}%
  \BibitemOpen
  \bibfield  {author} {\bibinfo {author} {\bibfnamefont {J.}~\bibnamefont
  {Walter}}, \bibinfo {author} {\bibfnamefont {S.}~\bibnamefont {Bose}},
  \bibinfo {author} {\bibfnamefont {M.}~\bibnamefont {Cabero}}, \bibinfo
  {author} {\bibfnamefont {M.}~\bibnamefont {Varela}},\ and\ \bibinfo {author}
  {\bibfnamefont {C.}~\bibnamefont {Leighton}},\ }\bibfield  {title} {\bibinfo
  {title} {Giant anisotropic magnetoresistance in oxygen-vacancy-ordered
  epitaxial {La}$_{0.5}${Sr}$_{0.5}${CoO}$_{3-\delta}$ films},\ }\href
  {https://doi.org/10.1103/PhysRevMaterials.4.091401} {\bibfield  {journal}
  {\bibinfo  {journal} {Phys. Rev. Mater.}\ }\textbf {\bibinfo {volume} {4}},\
  \bibinfo {pages} {091401} (\bibinfo {year} {2020})}\BibitemShut {NoStop}%
\bibitem [{\citenamefont {Tian}\ \emph {et~al.}(2018)\citenamefont {Tian},
  \citenamefont {Kuang}, \citenamefont {Mao}, \citenamefont {Yang},
  \citenamefont {Xiang}, \citenamefont {Xu}, \citenamefont {Sayedaghaee},
  \citenamefont {\'I\~niguez},\ and\ \citenamefont {Bellaiche}}]{Tian2018}%
  \BibitemOpen
  \bibfield  {author} {\bibinfo {author} {\bibfnamefont {H.}~\bibnamefont
  {Tian}}, \bibinfo {author} {\bibfnamefont {X.-Y.}\ \bibnamefont {Kuang}},
  \bibinfo {author} {\bibfnamefont {A.-J.}\ \bibnamefont {Mao}}, \bibinfo
  {author} {\bibfnamefont {Y.}~\bibnamefont {Yang}}, \bibinfo {author}
  {\bibfnamefont {H.}~\bibnamefont {Xiang}}, \bibinfo {author} {\bibfnamefont
  {C.}~\bibnamefont {Xu}}, \bibinfo {author} {\bibfnamefont {S.~O.}\
  \bibnamefont {Sayedaghaee}}, \bibinfo {author} {\bibfnamefont
  {J.}~\bibnamefont {\'I\~niguez}},\ and\ \bibinfo {author} {\bibfnamefont
  {L.}~\bibnamefont {Bellaiche}},\ }\bibfield  {title} {\bibinfo {title} {Novel
  type of ferroelectricity in brownmillerite structures: A first-principles
  study},\ }\href {https://doi.org/10.1103/PhysRevMaterials.2.084402}
  {\bibfield  {journal} {\bibinfo  {journal} {Phys. Rev. Materials}\ }\textbf
  {\bibinfo {volume} {2}},\ \bibinfo {pages} {084402} (\bibinfo {year}
  {2018})}\BibitemShut {NoStop}%
\bibitem [{\citenamefont {Young}\ and\ \citenamefont
  {Rondinelli}(2015)}]{Young2015}%
  \BibitemOpen
  \bibfield  {author} {\bibinfo {author} {\bibfnamefont {J.}~\bibnamefont
  {Young}}\ and\ \bibinfo {author} {\bibfnamefont {J.~M.}\ \bibnamefont
  {Rondinelli}},\ }\bibfield  {title} {\bibinfo {title} {{Crystal structure and
  electronic properties of bulk and thin film brownmillerite oxides}},\ }\href
  {https://doi.org/10.1103/PhysRevB.92.174111} {\bibfield  {journal} {\bibinfo
  {journal} {Physical Review B}\ }\textbf {\bibinfo {volume} {92}},\ \bibinfo
  {pages} {174111} (\bibinfo {year} {2015})}\BibitemShut {NoStop}%
\bibitem [{\citenamefont {Parsons}\ \emph {et~al.}(2009)\citenamefont
  {Parsons}, \citenamefont {D'Hondt}, \citenamefont {Hadermann},\ and\
  \citenamefont {Hayward}}]{Parsons2009}%
  \BibitemOpen
  \bibfield  {author} {\bibinfo {author} {\bibfnamefont {T.~G.}\ \bibnamefont
  {Parsons}}, \bibinfo {author} {\bibfnamefont {H.}~\bibnamefont {D'Hondt}},
  \bibinfo {author} {\bibfnamefont {J.}~\bibnamefont {Hadermann}},\ and\
  \bibinfo {author} {\bibfnamefont {M.~a.}\ \bibnamefont {Hayward}},\
  }\bibfield  {title} {\bibinfo {title} {{Synthesis and Structural
  Characterization of La$_{1-x}A_{x}$MnO$_{2.5}$ ($A$ = Ba, Sr, Ca) Phases:
  Mapping the Variants of the Brownmillerite Structure}},\ }\href
  {https://doi.org/10.1021/cm902535m} {\bibfield  {journal} {\bibinfo
  {journal} {Chemistry of Materials}\ }\textbf {\bibinfo {volume} {21}},\
  \bibinfo {pages} {5527} (\bibinfo {year} {2009})}\BibitemShut {NoStop}%
\bibitem [{\citenamefont {Marik}\ \emph {et~al.}(2019)\citenamefont {Marik},
  \citenamefont {Gonano}, \citenamefont {Veillon}, \citenamefont {Bréard},
  \citenamefont {Pelloquin}, \citenamefont {Hardy}, \citenamefont {Clet},\ and\
  \citenamefont {Le~Breton}}]{Marik/LeBreton2019}%
  \BibitemOpen
  \bibfield  {author} {\bibinfo {author} {\bibfnamefont {S.}~\bibnamefont
  {Marik}}, \bibinfo {author} {\bibfnamefont {B.}~\bibnamefont {Gonano}},
  \bibinfo {author} {\bibfnamefont {F.}~\bibnamefont {Veillon}}, \bibinfo
  {author} {\bibfnamefont {Y.}~\bibnamefont {Bréard}}, \bibinfo {author}
  {\bibfnamefont {D.}~\bibnamefont {Pelloquin}}, \bibinfo {author}
  {\bibfnamefont {V.}~\bibnamefont {Hardy}}, \bibinfo {author} {\bibfnamefont
  {G.}~\bibnamefont {Clet}},\ and\ \bibinfo {author} {\bibfnamefont {J.~M.}\
  \bibnamefont {Le~Breton}},\ }\bibfield  {title} {\bibinfo {title}
  {Tetrahedral chain ordering and low dimensional magnetic lattice in a new
  brownmillerite {Sr$_2$ScFeO$_5$}},\ }\href
  {https://doi.org/10.1039/C9CC05158A} {\bibfield  {journal} {\bibinfo
  {journal} {Chem. Commun.}\ }\textbf {\bibinfo {volume} {55}},\ \bibinfo
  {pages} {10436} (\bibinfo {year} {2019})}\BibitemShut {NoStop}%
\bibitem [{\citenamefont {Young}\ \emph {et~al.}(2017)\citenamefont {Young},
  \citenamefont {Moon}, \citenamefont {Mukherjee}, \citenamefont {Stone},
  \citenamefont {Gopalan}, \citenamefont {Alem}, \citenamefont {May},\ and\
  \citenamefont {Rondinelli}}]{Young2017}%
  \BibitemOpen
  \bibfield  {author} {\bibinfo {author} {\bibfnamefont {J.}~\bibnamefont
  {Young}}, \bibinfo {author} {\bibfnamefont {E.~J.}\ \bibnamefont {Moon}},
  \bibinfo {author} {\bibfnamefont {D.}~\bibnamefont {Mukherjee}}, \bibinfo
  {author} {\bibfnamefont {G.}~\bibnamefont {Stone}}, \bibinfo {author}
  {\bibfnamefont {V.}~\bibnamefont {Gopalan}}, \bibinfo {author} {\bibfnamefont
  {N.}~\bibnamefont {Alem}}, \bibinfo {author} {\bibfnamefont {S.~J.}\
  \bibnamefont {May}},\ and\ \bibinfo {author} {\bibfnamefont {J.~M.}\
  \bibnamefont {Rondinelli}},\ }\bibfield  {title} {\bibinfo {title} {Polar
  oxides without inversion symmetry through vacancy and chemical order},\
  }\href {https://doi.org/10.1021/jacs.6b10697} {\bibfield  {journal} {\bibinfo
   {journal} {Journal of the American Chemical Society}\ }\textbf {\bibinfo
  {volume} {139}},\ \bibinfo {pages} {2833} (\bibinfo {year}
  {2017})}\BibitemShut {NoStop}%
\bibitem [{\citenamefont {Kang}\ \emph {et~al.}(2019)\citenamefont {Kang},
  \citenamefont {Roh}, \citenamefont {Lim}, \citenamefont {Min}, \citenamefont
  {Lee}, \citenamefont {Lee}, \citenamefont {Lee}, \citenamefont {Kang},
  \citenamefont {Seol}, \citenamefont {Kim}, \citenamefont {Ohta},
  \citenamefont {Khare}, \citenamefont {Park}, \citenamefont {Kim},
  \citenamefont {Chae}, \citenamefont {Oh}, \citenamefont {Lee}, \citenamefont
  {Yu}, \citenamefont {Lee},\ and\ \citenamefont {Choi}}]{Kang/Choi2019}%
  \BibitemOpen
  \bibfield  {author} {\bibinfo {author} {\bibfnamefont {K.~T.}\ \bibnamefont
  {Kang}}, \bibinfo {author} {\bibfnamefont {C.~J.}\ \bibnamefont {Roh}},
  \bibinfo {author} {\bibfnamefont {J.}~\bibnamefont {Lim}}, \bibinfo {author}
  {\bibfnamefont {T.}~\bibnamefont {Min}}, \bibinfo {author} {\bibfnamefont
  {J.~H.}\ \bibnamefont {Lee}}, \bibinfo {author} {\bibfnamefont
  {K.}~\bibnamefont {Lee}}, \bibinfo {author} {\bibfnamefont {T.~Y.}\
  \bibnamefont {Lee}}, \bibinfo {author} {\bibfnamefont {S.}~\bibnamefont
  {Kang}}, \bibinfo {author} {\bibfnamefont {D.}~\bibnamefont {Seol}}, \bibinfo
  {author} {\bibfnamefont {J.}~\bibnamefont {Kim}}, \bibinfo {author}
  {\bibfnamefont {H.}~\bibnamefont {Ohta}}, \bibinfo {author} {\bibfnamefont
  {A.}~\bibnamefont {Khare}}, \bibinfo {author} {\bibfnamefont
  {S.}~\bibnamefont {Park}}, \bibinfo {author} {\bibfnamefont {Y.}~\bibnamefont
  {Kim}}, \bibinfo {author} {\bibfnamefont {S.~C.}\ \bibnamefont {Chae}},
  \bibinfo {author} {\bibfnamefont {Y.~S.}\ \bibnamefont {Oh}}, \bibinfo
  {author} {\bibfnamefont {J.}~\bibnamefont {Lee}}, \bibinfo {author}
  {\bibfnamefont {J.}~\bibnamefont {Yu}}, \bibinfo {author} {\bibfnamefont
  {J.~S.}\ \bibnamefont {Lee}},\ and\ \bibinfo {author} {\bibfnamefont {W.~S.}\
  \bibnamefont {Choi}},\ }\bibfield  {title} {\bibinfo {title} {A
  room-temperature ferroelectric ferromagnet in a {1D} tetrahedral chain
  network},\ }\href {https://doi.org/https://doi.org/10.1002/adma.201808104}
  {\bibfield  {journal} {\bibinfo  {journal} {Advanced Materials}\ }\textbf
  {\bibinfo {volume} {31}},\ \bibinfo {pages} {1808104} (\bibinfo {year}
  {2019})}\BibitemShut {NoStop}%
\bibitem [{\citenamefont {Arras}\ \emph {et~al.}(2023)\citenamefont {Arras},
  \citenamefont {Paillard},\ and\ \citenamefont
  {Bellaiche}}]{Arras/Bellaiche2023}%
  \BibitemOpen
  \bibfield  {author} {\bibinfo {author} {\bibfnamefont {R.}~\bibnamefont
  {Arras}}, \bibinfo {author} {\bibfnamefont {C.}~\bibnamefont {Paillard}},\
  and\ \bibinfo {author} {\bibfnamefont {L.}~\bibnamefont {Bellaiche}},\
  }\bibfield  {title} {\bibinfo {title} {Effect of an electric field on
  ferroelectric and piezoelectric properties of brownmillerite
  {Ca}$_{2}${Al}$_{2}${O}$_{5}$},\ }\href
  {https://doi.org/10.1103/PhysRevB.107.144107} {\bibfield  {journal} {\bibinfo
   {journal} {Phys. Rev. B}\ }\textbf {\bibinfo {volume} {107}},\ \bibinfo
  {pages} {144107} (\bibinfo {year} {2023})}\BibitemShut {NoStop}%
\bibitem [{\citenamefont {Grenier}\ \emph {et~al.}(1981)\citenamefont
  {Grenier}, \citenamefont {Pouchard},\ and\ \citenamefont
  {Hagenmuller}}]{Grenier1981}%
  \BibitemOpen
  \bibfield  {author} {\bibinfo {author} {\bibfnamefont {J.-C.}\ \bibnamefont
  {Grenier}}, \bibinfo {author} {\bibfnamefont {M.}~\bibnamefont {Pouchard}},\
  and\ \bibinfo {author} {\bibfnamefont {P.}~\bibnamefont {Hagenmuller}},\
  }\bibfield  {title} {\bibinfo {title} {Vacancy ordering in oxygen-deficient
  perovskite-related ferrites},\ }in\ \href@noop {} {\emph {\bibinfo
  {booktitle} {Ferrites {\textperiodcentered} Transitions Elements
  Luminescence}}}\ (\bibinfo  {publisher} {Springer Berlin Heidelberg},\
  \bibinfo {address} {Berlin, Heidelberg},\ \bibinfo {year} {1981})\ pp.\
  \bibinfo {pages} {1--25}\BibitemShut {NoStop}%
\bibitem [{\citenamefont {Yao}\ \emph {et~al.}(2017)\citenamefont {Yao},
  \citenamefont {Inkinen},\ and\ \citenamefont {van Dijken}}]{Yao/Dijken2017}%
  \BibitemOpen
  \bibfield  {author} {\bibinfo {author} {\bibfnamefont {L.}~\bibnamefont
  {Yao}}, \bibinfo {author} {\bibfnamefont {S.}~\bibnamefont {Inkinen}},\ and\
  \bibinfo {author} {\bibfnamefont {S.}~\bibnamefont {van Dijken}},\ }\bibfield
   {title} {\bibinfo {title} {Direct observation of oxygen vacancy-driven
  structural and resistive phase transitions in
  {La}$_{2/3}${Sr}$_{1/3}${MnO}$_3$},\ }\href
  {http://dx.doi.org/10.1038/ncomms14544} {\bibfield  {journal} {\bibinfo
  {journal} {Nature Communications}\ }\textbf {\bibinfo {volume} {8}},\
  \bibinfo {pages} {14544} (\bibinfo {year} {2017})}\BibitemShut {NoStop}%
\bibitem [{\citenamefont {Luo}\ and\ \citenamefont
  {Hayward}(2013)}]{Luo/Hayward2013}%
  \BibitemOpen
  \bibfield  {author} {\bibinfo {author} {\bibfnamefont {K.}~\bibnamefont
  {Luo}}\ and\ \bibinfo {author} {\bibfnamefont {M.~A.}\ \bibnamefont
  {Hayward}},\ }\bibfield  {title} {\bibinfo {title} {{The synthesis and
  characterisation of LaCa$_2$Fe$_2$GaO$_8$}},\ }\href
  {https://doi.org/https://doi.org/10.1016/j.jssc.2012.10.012} {\bibfield
  {journal} {\bibinfo  {journal} {Journal of Solid State Chemistry}\ }\textbf
  {\bibinfo {volume} {198}},\ \bibinfo {pages} {203} (\bibinfo {year}
  {2013})}\BibitemShut {NoStop}%
\bibitem [{\citenamefont {Zhang}\ and\ \citenamefont
  {Galli}(2020)}]{Zhang/Galli2020}%
  \BibitemOpen
  \bibfield  {author} {\bibinfo {author} {\bibfnamefont {S.}~\bibnamefont
  {Zhang}}\ and\ \bibinfo {author} {\bibfnamefont {G.}~\bibnamefont {Galli}},\
  }\bibfield  {title} {\bibinfo {title} {{Understanding the metal-to-insulator
  transition in La$_{1-x}$Sr$_x$CoO$_{3-\delta}$ and its applications for
  neuromorphic computing}},\ }\href
  {https://www.nature.com/articles/s41524-020-00437-w} {\bibfield  {journal}
  {\bibinfo  {journal} {npj Computational Materials}\ }\textbf {\bibinfo
  {volume} {6}} (\bibinfo {year} {2020})}\BibitemShut {NoStop}%
\bibitem [{\citenamefont {Lu}\ \emph {et~al.}(2017)\citenamefont {Lu},
  \citenamefont {Zhang}, \citenamefont {Zhang}, \citenamefont {Qiao},
  \citenamefont {He}, \citenamefont {Li}, \citenamefont {Wang}, \citenamefont
  {Guo}, \citenamefont {Zhang}, \citenamefont {Duan}, \citenamefont {Li},
  \citenamefont {Wang}, \citenamefont {Yang}, \citenamefont {Yan},
  \citenamefont {Arenholz}, \citenamefont {Zhou}, \citenamefont {Yang},
  \citenamefont {Gu}, \citenamefont {Nan}, \citenamefont {Wu}, \citenamefont
  {Tokura},\ and\ \citenamefont {Yu}}]{Lu2017}%
  \BibitemOpen
  \bibfield  {author} {\bibinfo {author} {\bibfnamefont {N.}~\bibnamefont
  {Lu}}, \bibinfo {author} {\bibfnamefont {P.}~\bibnamefont {Zhang}}, \bibinfo
  {author} {\bibfnamefont {Q.}~\bibnamefont {Zhang}}, \bibinfo {author}
  {\bibfnamefont {R.}~\bibnamefont {Qiao}}, \bibinfo {author} {\bibfnamefont
  {Q.}~\bibnamefont {He}}, \bibinfo {author} {\bibfnamefont {H.-B.}\
  \bibnamefont {Li}}, \bibinfo {author} {\bibfnamefont {Y.}~\bibnamefont
  {Wang}}, \bibinfo {author} {\bibfnamefont {J.}~\bibnamefont {Guo}}, \bibinfo
  {author} {\bibfnamefont {D.}~\bibnamefont {Zhang}}, \bibinfo {author}
  {\bibfnamefont {Z.}~\bibnamefont {Duan}}, \bibinfo {author} {\bibfnamefont
  {Z.}~\bibnamefont {Li}}, \bibinfo {author} {\bibfnamefont {M.}~\bibnamefont
  {Wang}}, \bibinfo {author} {\bibfnamefont {S.}~\bibnamefont {Yang}}, \bibinfo
  {author} {\bibfnamefont {M.}~\bibnamefont {Yan}}, \bibinfo {author}
  {\bibfnamefont {E.}~\bibnamefont {Arenholz}}, \bibinfo {author}
  {\bibfnamefont {S.}~\bibnamefont {Zhou}}, \bibinfo {author} {\bibfnamefont
  {W.}~\bibnamefont {Yang}}, \bibinfo {author} {\bibfnamefont {L.}~\bibnamefont
  {Gu}}, \bibinfo {author} {\bibfnamefont {C.-W.}\ \bibnamefont {Nan}},
  \bibinfo {author} {\bibfnamefont {J.}~\bibnamefont {Wu}}, \bibinfo {author}
  {\bibfnamefont {Y.}~\bibnamefont {Tokura}},\ and\ \bibinfo {author}
  {\bibfnamefont {P.}~\bibnamefont {Yu}},\ }\bibfield  {title} {\bibinfo
  {title} {Electric-field control of tri-state phase transformation with a
  selective dual-ion switch},\ }\href {http://dx.doi.org/10.1038/nature22389}
  {\bibfield  {journal} {\bibinfo  {journal} {Nature}\ }\textbf {\bibinfo
  {volume} {546}},\ \bibinfo {pages} {124} (\bibinfo {year}
  {2017})}\BibitemShut {NoStop}%
\bibitem [{\citenamefont {Shaula}\ \emph {et~al.}(2006)\citenamefont {Shaula},
  \citenamefont {Pivak}, \citenamefont {Waerenborgh}, \citenamefont
  {Gaczyñski}, \citenamefont {Yaremchenko},\ and\ \citenamefont
  {Kharton}}]{Shaula2006}%
  \BibitemOpen
  \bibfield  {author} {\bibinfo {author} {\bibfnamefont {A.}~\bibnamefont
  {Shaula}}, \bibinfo {author} {\bibfnamefont {Y.}~\bibnamefont {Pivak}},
  \bibinfo {author} {\bibfnamefont {J.}~\bibnamefont {Waerenborgh}}, \bibinfo
  {author} {\bibfnamefont {P.}~\bibnamefont {Gaczyñski}}, \bibinfo {author}
  {\bibfnamefont {A.}~\bibnamefont {Yaremchenko}},\ and\ \bibinfo {author}
  {\bibfnamefont {V.}~\bibnamefont {Kharton}},\ }\bibfield  {title} {\bibinfo
  {title} {Ionic conductivity of brownmillerite-type calcium ferrite under
  oxidizing conditions},\ }\href
  {https://doi.org/https://doi.org/10.1016/j.ssi.2006.08.030} {\bibfield
  {journal} {\bibinfo  {journal} {Solid State Ionics}\ }\textbf {\bibinfo
  {volume} {177}},\ \bibinfo {pages} {2923} (\bibinfo {year}
  {2006})}\BibitemShut {NoStop}%
\bibitem [{\citenamefont {Orera}\ and\ \citenamefont
  {Slater}(2010)}]{Orera2010}%
  \BibitemOpen
  \bibfield  {author} {\bibinfo {author} {\bibfnamefont {A.}~\bibnamefont
  {Orera}}\ and\ \bibinfo {author} {\bibfnamefont {P.~R.}\ \bibnamefont
  {Slater}},\ }\bibfield  {title} {\bibinfo {title} {New chemical systems for
  solid oxide fuel cells},\ }\href {https://doi.org/10.1021/cm902687z}
  {\bibfield  {journal} {\bibinfo  {journal} {Chemistry of Materials}\ }\textbf
  {\bibinfo {volume} {22}},\ \bibinfo {pages} {675} (\bibinfo {year}
  {2010})}\BibitemShut {NoStop}%
\bibitem [{\citenamefont {Lu}\ \emph {et~al.}(2022)\citenamefont {Lu},
  \citenamefont {Zhang}, \citenamefont {Wang}, \citenamefont {Li},
  \citenamefont {Qiao}, \citenamefont {Zhao}, \citenamefont {He}, \citenamefont
  {Lu}, \citenamefont {Li}, \citenamefont {Wu}, \citenamefont {Zhu},
  \citenamefont {Lyu}, \citenamefont {Chen}, \citenamefont {Li}, \citenamefont
  {Wang}, \citenamefont {Zhang}, \citenamefont {Tsang}, \citenamefont {Guo},
  \citenamefont {Yang}, \citenamefont {Zhang}, \citenamefont {Deng},
  \citenamefont {Zhang}, \citenamefont {Ma}, \citenamefont {Ren}, \citenamefont
  {Wu}, \citenamefont {Zhu}, \citenamefont {Zhou}, \citenamefont {Tokura},
  \citenamefont {Nan}, \citenamefont {Wu},\ and\ \citenamefont
  {Yu}}]{PuYu2022}%
  \BibitemOpen
  \bibfield  {author} {\bibinfo {author} {\bibfnamefont {N.}~\bibnamefont
  {Lu}}, \bibinfo {author} {\bibfnamefont {Z.}~\bibnamefont {Zhang}}, \bibinfo
  {author} {\bibfnamefont {Y.}~\bibnamefont {Wang}}, \bibinfo {author}
  {\bibfnamefont {H.-B.}\ \bibnamefont {Li}}, \bibinfo {author} {\bibfnamefont
  {S.}~\bibnamefont {Qiao}}, \bibinfo {author} {\bibfnamefont {B.}~\bibnamefont
  {Zhao}}, \bibinfo {author} {\bibfnamefont {Q.}~\bibnamefont {He}}, \bibinfo
  {author} {\bibfnamefont {S.}~\bibnamefont {Lu}}, \bibinfo {author}
  {\bibfnamefont {C.}~\bibnamefont {Li}}, \bibinfo {author} {\bibfnamefont
  {Y.}~\bibnamefont {Wu}}, \bibinfo {author} {\bibfnamefont {M.}~\bibnamefont
  {Zhu}}, \bibinfo {author} {\bibfnamefont {X.}~\bibnamefont {Lyu}}, \bibinfo
  {author} {\bibfnamefont {X.}~\bibnamefont {Chen}}, \bibinfo {author}
  {\bibfnamefont {Z.}~\bibnamefont {Li}}, \bibinfo {author} {\bibfnamefont
  {M.}~\bibnamefont {Wang}}, \bibinfo {author} {\bibfnamefont {J.}~\bibnamefont
  {Zhang}}, \bibinfo {author} {\bibfnamefont {S.~C.}\ \bibnamefont {Tsang}},
  \bibinfo {author} {\bibfnamefont {J.}~\bibnamefont {Guo}}, \bibinfo {author}
  {\bibfnamefont {S.}~\bibnamefont {Yang}}, \bibinfo {author} {\bibfnamefont
  {J.}~\bibnamefont {Zhang}}, \bibinfo {author} {\bibfnamefont
  {K.}~\bibnamefont {Deng}}, \bibinfo {author} {\bibfnamefont {D.}~\bibnamefont
  {Zhang}}, \bibinfo {author} {\bibfnamefont {J.}~\bibnamefont {Ma}}, \bibinfo
  {author} {\bibfnamefont {J.}~\bibnamefont {Ren}}, \bibinfo {author}
  {\bibfnamefont {Y.}~\bibnamefont {Wu}}, \bibinfo {author} {\bibfnamefont
  {J.}~\bibnamefont {Zhu}}, \bibinfo {author} {\bibfnamefont {S.}~\bibnamefont
  {Zhou}}, \bibinfo {author} {\bibfnamefont {Y.}~\bibnamefont {Tokura}},
  \bibinfo {author} {\bibfnamefont {C.-W.}\ \bibnamefont {Nan}}, \bibinfo
  {author} {\bibfnamefont {J.}~\bibnamefont {Wu}},\ and\ \bibinfo {author}
  {\bibfnamefont {P.}~\bibnamefont {Yu}},\ }\bibfield  {title} {\bibinfo
  {title} {Enhanced low-temperature proton conductivity in
  hydrogen-intercalated brownmillerite oxide},\ }\href
  {https://doi.org/10.1038/s41560-022-01166-8} {\bibfield  {journal} {\bibinfo
  {journal} {Nature Energy}\ }\textbf {\bibinfo {volume} {7}},\ \bibinfo
  {pages} {1208} (\bibinfo {year} {2022})}\BibitemShut {NoStop}%
\bibitem [{\citenamefont {Karki}\ and\ \citenamefont
  {Ramezanipour}(2020)}]{Karki2020}%
  \BibitemOpen
  \bibfield  {author} {\bibinfo {author} {\bibfnamefont {S.~B.}\ \bibnamefont
  {Karki}}\ and\ \bibinfo {author} {\bibfnamefont {F.}~\bibnamefont
  {Ramezanipour}},\ }\bibfield  {title} {\bibinfo {title} {Pseudocapacitive
  energy storage and electrocatalytic hydrogen-evolution activity of
  defect-ordered perovskites {Sr}$_x${Ca}$_{3–x}${GaMn}$_2${O}$_8$ ($x$ = 0
  and 1)},\ }\href {https://doi.org/10.1021/acsaem.0c01935} {\bibfield
  {journal} {\bibinfo  {journal} {ACS Applied Energy Materials}\ }\textbf
  {\bibinfo {volume} {3}},\ \bibinfo {pages} {10983} (\bibinfo {year}
  {2020})}\BibitemShut {NoStop}%
\bibitem [{\citenamefont {Xie}\ \emph {et~al.}(2014)\citenamefont {Xie},
  \citenamefont {Scafetta}, \citenamefont {Sichel-Tissot}, \citenamefont
  {Moon}, \citenamefont {Devlin}, \citenamefont {Wu}, \citenamefont {Krick},\
  and\ \citenamefont {May}}]{May2014}%
  \BibitemOpen
  \bibfield  {author} {\bibinfo {author} {\bibfnamefont {Y.}~\bibnamefont
  {Xie}}, \bibinfo {author} {\bibfnamefont {M.~D.}\ \bibnamefont {Scafetta}},
  \bibinfo {author} {\bibfnamefont {R.~J.}\ \bibnamefont {Sichel-Tissot}},
  \bibinfo {author} {\bibfnamefont {E.~J.}\ \bibnamefont {Moon}}, \bibinfo
  {author} {\bibfnamefont {R.~C.}\ \bibnamefont {Devlin}}, \bibinfo {author}
  {\bibfnamefont {H.}~\bibnamefont {Wu}}, \bibinfo {author} {\bibfnamefont
  {A.~L.}\ \bibnamefont {Krick}},\ and\ \bibinfo {author} {\bibfnamefont
  {S.~J.}\ \bibnamefont {May}},\ }\bibfield  {title} {\bibinfo {title}
  {{Control of Functional Responses Via Reversible Oxygen Loss in
  La$_{1-x}$Sr$_x$FeO$_{3-\delta}$ Films}},\ }\href
  {https://doi.org/https://doi.org/10.1002/adma.201304374} {\bibfield
  {journal} {\bibinfo  {journal} {Advanced Materials}\ }\textbf {\bibinfo
  {volume} {26}},\ \bibinfo {pages} {1434} (\bibinfo {year}
  {2014})}\BibitemShut {NoStop}%
\bibitem [{\citenamefont {Islam}\ \emph {et~al.}(2015)\citenamefont {Islam},
  \citenamefont {Xie}, \citenamefont {Scafetta}, \citenamefont {May},\ and\
  \citenamefont {Spanier}}]{May/Spanier2015}%
  \BibitemOpen
  \bibfield  {author} {\bibinfo {author} {\bibfnamefont {M.~A.}\ \bibnamefont
  {Islam}}, \bibinfo {author} {\bibfnamefont {Y.}~\bibnamefont {Xie}}, \bibinfo
  {author} {\bibfnamefont {M.~D.}\ \bibnamefont {Scafetta}}, \bibinfo {author}
  {\bibfnamefont {S.~J.}\ \bibnamefont {May}},\ and\ \bibinfo {author}
  {\bibfnamefont {J.~E.}\ \bibnamefont {Spanier}},\ }\bibfield  {title}
  {\bibinfo {title} {{Raman scattering in La$_{1-x}$Sr$_x$FeO$_{3-\delta}$ thin
  films: annealing-induced reduction and phase transformation}},\ }\href
  {https://doi.org/10.1088/0953-8984/27/15/155401} {\bibfield  {journal}
  {\bibinfo  {journal} {Journal of Physics: Condensed Matter}\ }\textbf
  {\bibinfo {volume} {27}},\ \bibinfo {pages} {155401} (\bibinfo {year}
  {2015})}\BibitemShut {NoStop}%
\bibitem [{\citenamefont {Battle}\ \emph {et~al.}(1990)\citenamefont {Battle},
  \citenamefont {Gibb},\ and\ \citenamefont {Lightfoot}}]{Battle1990}%
  \BibitemOpen
  \bibfield  {author} {\bibinfo {author} {\bibfnamefont {P.}~\bibnamefont
  {Battle}}, \bibinfo {author} {\bibfnamefont {T.}~\bibnamefont {Gibb}},\ and\
  \bibinfo {author} {\bibfnamefont {P.}~\bibnamefont {Lightfoot}},\ }\bibfield
  {title} {\bibinfo {title} {{The crystal and magnetic structures of
  Sr$_2$LaFe$_3$O$_8$}},\ }\href
  {https://doi.org/https://doi.org/10.1016/0022-4596(90)90322-O} {\bibfield
  {journal} {\bibinfo  {journal} {Journal of Solid State Chemistry}\ }\textbf
  {\bibinfo {volume} {84}},\ \bibinfo {pages} {237} (\bibinfo {year}
  {1990})}\BibitemShut {NoStop}%
\bibitem [{\citenamefont {Smolin}\ \emph {et~al.}(2016)\citenamefont {Smolin},
  \citenamefont {Scafetta}, \citenamefont {Choquette}, \citenamefont {Sfeir},
  \citenamefont {Baxter},\ and\ \citenamefont {May}}]{May2016}%
  \BibitemOpen
  \bibfield  {author} {\bibinfo {author} {\bibfnamefont {S.~Y.}\ \bibnamefont
  {Smolin}}, \bibinfo {author} {\bibfnamefont {M.~D.}\ \bibnamefont
  {Scafetta}}, \bibinfo {author} {\bibfnamefont {A.~K.}\ \bibnamefont
  {Choquette}}, \bibinfo {author} {\bibfnamefont {M.~Y.}\ \bibnamefont
  {Sfeir}}, \bibinfo {author} {\bibfnamefont {J.~B.}\ \bibnamefont {Baxter}},\
  and\ \bibinfo {author} {\bibfnamefont {S.~J.}\ \bibnamefont {May}},\
  }\bibfield  {title} {\bibinfo {title} {{Static and Dynamic Optical Properties
  of La$_{1-x}$Sr$_x$FeO$_{3-\delta}$: The Effects of $A$-Site and Oxygen
  Stoichiometry}},\ }\href {https://doi.org/10.1021/acs.chemmater.5b03273}
  {\bibfield  {journal} {\bibinfo  {journal} {Chemistry of Materials}\ }\textbf
  {\bibinfo {volume} {28}},\ \bibinfo {pages} {97} (\bibinfo {year}
  {2016})}\BibitemShut {NoStop}%
\bibitem [{\citenamefont {Guo}\ \emph {et~al.}(2017)\citenamefont {Guo},
  \citenamefont {Hosaka}, \citenamefont {Romero}, \citenamefont {Saito},
  \citenamefont {Ichikawa},\ and\ \citenamefont {Shimakawa}}]{Shimakawa2017}%
  \BibitemOpen
  \bibfield  {author} {\bibinfo {author} {\bibfnamefont {H.}~\bibnamefont
  {Guo}}, \bibinfo {author} {\bibfnamefont {Y.}~\bibnamefont {Hosaka}},
  \bibinfo {author} {\bibfnamefont {F.~D.}\ \bibnamefont {Romero}}, \bibinfo
  {author} {\bibfnamefont {T.}~\bibnamefont {Saito}}, \bibinfo {author}
  {\bibfnamefont {N.}~\bibnamefont {Ichikawa}},\ and\ \bibinfo {author}
  {\bibfnamefont {Y.}~\bibnamefont {Shimakawa}},\ }\bibfield  {title} {\bibinfo
  {title} {{Two Charge Ordering Patterns in the Topochemically Synthesized
  Layer-Structured Perovskite LaCa$_2$Fe$_3$O$_9$ with Unusually High Valence
  Fe$^{3.67+}$}},\ }\href {https://doi.org/10.1021/acs.inorgchem.7b00104}
  {\bibfield  {journal} {\bibinfo  {journal} {Inorganic Chemistry}\ }\textbf
  {\bibinfo {volume} {56}},\ \bibinfo {pages} {3695} (\bibinfo {year}
  {2017})}\BibitemShut {NoStop}%
\bibitem [{\citenamefont {Guo}\ \emph {et~al.}(2022)\citenamefont {Guo},
  \citenamefont {Patino}, \citenamefont {Ichikawa}, \citenamefont {Saito},
  \citenamefont {Watanabe}, \citenamefont {Goto}, \citenamefont {Yang},
  \citenamefont {Kan},\ and\ \citenamefont {Shimakawa}}]{Shimakawa2022}%
  \BibitemOpen
  \bibfield  {author} {\bibinfo {author} {\bibfnamefont {H.}~\bibnamefont
  {Guo}}, \bibinfo {author} {\bibfnamefont {M.~A.}\ \bibnamefont {Patino}},
  \bibinfo {author} {\bibfnamefont {N.}~\bibnamefont {Ichikawa}}, \bibinfo
  {author} {\bibfnamefont {T.}~\bibnamefont {Saito}}, \bibinfo {author}
  {\bibfnamefont {R.}~\bibnamefont {Watanabe}}, \bibinfo {author}
  {\bibfnamefont {M.}~\bibnamefont {Goto}}, \bibinfo {author} {\bibfnamefont
  {M.}~\bibnamefont {Yang}}, \bibinfo {author} {\bibfnamefont {D.}~\bibnamefont
  {Kan}},\ and\ \bibinfo {author} {\bibfnamefont {Y.}~\bibnamefont
  {Shimakawa}},\ }\bibfield  {title} {\bibinfo {title} {{Oxygen Release and
  Incorporation Behaviors Influenced by $A$-Site Cation Order/Disorder in
  LaCa$_2$Fe$_3$O$_9$ with Unusually High Valence Fe$^{3.67+}$}},\ }\href
  {https://doi.org/10.1021/acs.chemmater.1c03686} {\bibfield  {journal}
  {\bibinfo  {journal} {Chemistry of Materials}\ }\textbf {\bibinfo {volume}
  {34}},\ \bibinfo {pages} {345} (\bibinfo {year} {2022})}\BibitemShut
  {NoStop}%
\bibitem [{\citenamefont {Hudspeth}\ \emph {et~al.}(2009)\citenamefont
  {Hudspeth}, \citenamefont {Goossens}, \citenamefont {Studer}, \citenamefont
  {Withers},\ and\ \citenamefont {Norén}}]{Hudspeth2009}%
  \BibitemOpen
  \bibfield  {author} {\bibinfo {author} {\bibfnamefont {J.~M.}\ \bibnamefont
  {Hudspeth}}, \bibinfo {author} {\bibfnamefont {D.~J.}\ \bibnamefont
  {Goossens}}, \bibinfo {author} {\bibfnamefont {A.~J.}\ \bibnamefont
  {Studer}}, \bibinfo {author} {\bibfnamefont {R.~L.}\ \bibnamefont
  {Withers}},\ and\ \bibinfo {author} {\bibfnamefont {L.}~\bibnamefont
  {Norén}},\ }\bibfield  {title} {\bibinfo {title} {{The crystal and magnetic
  structures of LaCa$_2$Fe$_3$O$_8$ and NdCa$_2$Fe$_3$O$_8$}},\ }\href
  {https://doi.org/10.1088/0953-8984/21/12/124206} {\bibfield  {journal}
  {\bibinfo  {journal} {Journal of Physics: Condensed Matter}\ }\textbf
  {\bibinfo {volume} {21}},\ \bibinfo {pages} {124206} (\bibinfo {year}
  {2009})}\BibitemShut {NoStop}%
\bibitem [{\citenamefont {Mart\'{i}nez~de Irujo-Labalde}\ \emph
  {et~al.}(2019)\citenamefont {Mart\'{i}nez~de Irujo-Labalde}, \citenamefont
  {Goto}, \citenamefont {Urones-Garrote}, \citenamefont {Amador}, \citenamefont
  {Ritter}, \citenamefont {Amano~Patino}, \citenamefont {Koedtruad},
  \citenamefont {Tan}, \citenamefont {Shimakawa},\ and\ \citenamefont
  {García-Martín}}]{Shimakawa/Martin2019}%
  \BibitemOpen
  \bibfield  {author} {\bibinfo {author} {\bibfnamefont {X.}~\bibnamefont
  {Mart\'{i}nez~de Irujo-Labalde}}, \bibinfo {author} {\bibfnamefont
  {M.}~\bibnamefont {Goto}}, \bibinfo {author} {\bibfnamefont {E.}~\bibnamefont
  {Urones-Garrote}}, \bibinfo {author} {\bibfnamefont {U.}~\bibnamefont
  {Amador}}, \bibinfo {author} {\bibfnamefont {C.}~\bibnamefont {Ritter}},
  \bibinfo {author} {\bibfnamefont {M.~E.}\ \bibnamefont {Amano~Patino}},
  \bibinfo {author} {\bibfnamefont {A.}~\bibnamefont {Koedtruad}}, \bibinfo
  {author} {\bibfnamefont {Z.}~\bibnamefont {Tan}}, \bibinfo {author}
  {\bibfnamefont {Y.}~\bibnamefont {Shimakawa}},\ and\ \bibinfo {author}
  {\bibfnamefont {S.}~\bibnamefont {García-Martín}},\ }\bibfield  {title}
  {\bibinfo {title} {Multiferroism induced by spontaneous structural ordering
  in antiferromagnetic iron perovskites},\ }\href
  {https://doi.org/10.1021/acs.chemmater.9b02716} {\bibfield  {journal}
  {\bibinfo  {journal} {Chemistry of Materials}\ }\textbf {\bibinfo {volume}
  {31}},\ \bibinfo {pages} {5993} (\bibinfo {year} {2019})}\BibitemShut
  {NoStop}%
\bibitem [{\citenamefont {Khan}\ \emph {et~al.}(2020)\citenamefont {Khan},
  \citenamefont {Keshavarzi},\ and\ \citenamefont {Datta}}]{Khan2020}%
  \BibitemOpen
  \bibfield  {author} {\bibinfo {author} {\bibfnamefont {A.~I.}\ \bibnamefont
  {Khan}}, \bibinfo {author} {\bibfnamefont {A.}~\bibnamefont {Keshavarzi}},\
  and\ \bibinfo {author} {\bibfnamefont {S.}~\bibnamefont {Datta}},\ }\bibfield
   {title} {\bibinfo {title} {The future of ferroelectric field-effect
  transistor technology},\ }\href {https://doi.org/10.1038/s41928-020-00492-7}
  {\bibfield  {journal} {\bibinfo  {journal} {Nature Electronics}\ }\textbf
  {\bibinfo {volume} {3}},\ \bibinfo {pages} {588} (\bibinfo {year}
  {2020})}\BibitemShut {NoStop}%
\bibitem [{\citenamefont {Zhang}\ \emph {et~al.}(2021)\citenamefont {Zhang},
  \citenamefont {Białek}, \citenamefont {Magrez}, \citenamefont {Yu},\ and\
  \citenamefont {Ansermet}}]{Zhang/Ansermet2021}%
  \BibitemOpen
  \bibfield  {author} {\bibinfo {author} {\bibfnamefont {J.}~\bibnamefont
  {Zhang}}, \bibinfo {author} {\bibfnamefont {M.}~\bibnamefont {Białek}},
  \bibinfo {author} {\bibfnamefont {A.}~\bibnamefont {Magrez}}, \bibinfo
  {author} {\bibfnamefont {H.}~\bibnamefont {Yu}},\ and\ \bibinfo {author}
  {\bibfnamefont {J.-P.}\ \bibnamefont {Ansermet}},\ }\bibfield  {title}
  {\bibinfo {title} {{Antiferromagnetic resonance in TmFeO$_3$ at high
  temperatures}},\ }\href
  {https://doi.org/https://doi.org/10.1016/j.jmmm.2020.167562} {\bibfield
  {journal} {\bibinfo  {journal} {Journal of Magnetism and Magnetic Materials}\
  }\textbf {\bibinfo {volume} {523}},\ \bibinfo {pages} {167562} (\bibinfo
  {year} {2021})}\BibitemShut {NoStop}%
\bibitem [{\citenamefont {Aparnadevi}\ \emph {et~al.}(2016)\citenamefont
  {Aparnadevi}, \citenamefont {Saravana~Kumar}, \citenamefont {Manikandan},
  \citenamefont {Paul~Joseph},\ and\ \citenamefont
  {Venkateswaran}}]{Aparnadevi2016}%
  \BibitemOpen
  \bibfield  {author} {\bibinfo {author} {\bibfnamefont {N.}~\bibnamefont
  {Aparnadevi}}, \bibinfo {author} {\bibfnamefont {K.}~\bibnamefont
  {Saravana~Kumar}}, \bibinfo {author} {\bibfnamefont {M.}~\bibnamefont
  {Manikandan}}, \bibinfo {author} {\bibfnamefont {D.}~\bibnamefont
  {Paul~Joseph}},\ and\ \bibinfo {author} {\bibfnamefont {C.}~\bibnamefont
  {Venkateswaran}},\ }\bibfield  {title} {\bibinfo {title} {{Room temperature
  dual ferroic behaviour of ball mill synthesized NdFeO$_3$ orthoferrite}},\
  }\href {https://doi.org/10.1063/1.4954842} {\bibfield  {journal} {\bibinfo
  {journal} {Journal of Applied Physics}\ }\textbf {\bibinfo {volume} {120}},\
  \bibinfo {pages} {034101} (\bibinfo {year} {2016})}\BibitemShut {NoStop}%
\bibitem [{\citenamefont {S{\l}awi{\'{n}}ski}\ \emph
  {et~al.}(2005)\citenamefont {S{\l}awi{\'{n}}ski}, \citenamefont
  {Przenios{\l}o}, \citenamefont {Sosnowska},\ and\ \citenamefont
  {Suard}}]{Slawinski2005}%
  \BibitemOpen
  \bibfield  {author} {\bibinfo {author} {\bibfnamefont {W.}~\bibnamefont
  {S{\l}awi{\'{n}}ski}}, \bibinfo {author} {\bibfnamefont {R.}~\bibnamefont
  {Przenios{\l}o}}, \bibinfo {author} {\bibfnamefont {I.}~\bibnamefont
  {Sosnowska}},\ and\ \bibinfo {author} {\bibfnamefont {E.}~\bibnamefont
  {Suard}},\ }\bibfield  {title} {\bibinfo {title} {Spin reorientation and
  structural changes in {NdFeO}$_3$},\ }\href
  {https://doi.org/10.1088/0953-8984/17/29/002} {\bibfield  {journal} {\bibinfo
   {journal} {Journal of Physics: Condensed Matter}\ }\textbf {\bibinfo
  {volume} {17}},\ \bibinfo {pages} {4605} (\bibinfo {year}
  {2005})}\BibitemShut {NoStop}%
\bibitem [{\citenamefont {Ritter}\ \emph {et~al.}(2022)\citenamefont {Ritter},
  \citenamefont {Vilarinho}, \citenamefont {Moreira}, \citenamefont {Mihalik},
  \citenamefont {Mihalik},\ and\ \citenamefont {Savvin}}]{Ritter2022}%
  \BibitemOpen
  \bibfield  {author} {\bibinfo {author} {\bibfnamefont {C.}~\bibnamefont
  {Ritter}}, \bibinfo {author} {\bibfnamefont {R.}~\bibnamefont {Vilarinho}},
  \bibinfo {author} {\bibfnamefont {J.~A.}\ \bibnamefont {Moreira}}, \bibinfo
  {author} {\bibfnamefont {M.}~\bibnamefont {Mihalik}}, \bibinfo {author}
  {\bibfnamefont {M.}~\bibnamefont {Mihalik}},\ and\ \bibinfo {author}
  {\bibfnamefont {S.}~\bibnamefont {Savvin}},\ }\bibfield  {title} {\bibinfo
  {title} {{The magnetic structure of {DyFeO}$_3$ revisited: Fe spin
  reorientation and Dy incommensurate magnetic order}},\ }\href
  {https://doi.org/10.1088/1361-648x/ac6787} {\bibfield  {journal} {\bibinfo
  {journal} {Journal of Physics: Condensed Matter}\ }\textbf {\bibinfo {volume}
  {34}},\ \bibinfo {pages} {265801} (\bibinfo {year} {2022})}\BibitemShut
  {NoStop}%
\bibitem [{\citenamefont {Ritter}\ \emph {et~al.}(2021)\citenamefont {Ritter},
  \citenamefont {Ceretti},\ and\ \citenamefont {Paulus}}]{Ritter2021}%
  \BibitemOpen
  \bibfield  {author} {\bibinfo {author} {\bibfnamefont {C.}~\bibnamefont
  {Ritter}}, \bibinfo {author} {\bibfnamefont {M.}~\bibnamefont {Ceretti}},\
  and\ \bibinfo {author} {\bibfnamefont {W.}~\bibnamefont {Paulus}},\
  }\bibfield  {title} {\bibinfo {title} {Determination of the magnetic
  structures in orthoferrite {CeFeO}$_3$ by neutron powder diffraction: first
  order spin reorientation and appearance of an ordered {Ce}-moment},\ }\href
  {https://doi.org/10.1088/1361-648x/abe64a} {\bibfield  {journal} {\bibinfo
  {journal} {Journal of Physics: Condensed Matter}\ }\textbf {\bibinfo {volume}
  {33}},\ \bibinfo {pages} {215802} (\bibinfo {year} {2021})}\BibitemShut
  {NoStop}%
\bibitem [{\citenamefont {Prakash}\ \emph {et~al.}(2019)\citenamefont
  {Prakash}, \citenamefont {Sathe}, \citenamefont {Prajapat}, \citenamefont
  {Nigam}, \citenamefont {Krishna},\ and\ \citenamefont {Das}}]{Prakash2019}%
  \BibitemOpen
  \bibfield  {author} {\bibinfo {author} {\bibfnamefont {P.}~\bibnamefont
  {Prakash}}, \bibinfo {author} {\bibfnamefont {V.}~\bibnamefont {Sathe}},
  \bibinfo {author} {\bibfnamefont {C.~L.}\ \bibnamefont {Prajapat}}, \bibinfo
  {author} {\bibfnamefont {A.~K.}\ \bibnamefont {Nigam}}, \bibinfo {author}
  {\bibfnamefont {P.~S.~R.}\ \bibnamefont {Krishna}},\ and\ \bibinfo {author}
  {\bibfnamefont {A.}~\bibnamefont {Das}},\ }\bibfield  {title} {\bibinfo
  {title} {Spin phonon coupling in mn doped {HoFeO}$_3$ compounds exhibiting
  spin reorientation behaviour},\ }\href
  {https://doi.org/10.1088/1361-648x/ab576d} {\bibfield  {journal} {\bibinfo
  {journal} {Journal of Physics: Condensed Matter}\ }\textbf {\bibinfo {volume}
  {32}},\ \bibinfo {pages} {095801} (\bibinfo {year} {2019})}\BibitemShut
  {NoStop}%
\bibitem [{\citenamefont {Cao}\ \emph {et~al.}(2014)\citenamefont {Cao},
  \citenamefont {Zhao}, \citenamefont {Kang}, \citenamefont {Zhang},\ and\
  \citenamefont {Ren}}]{Cao2014}%
  \BibitemOpen
  \bibfield  {author} {\bibinfo {author} {\bibfnamefont {S.}~\bibnamefont
  {Cao}}, \bibinfo {author} {\bibfnamefont {H.}~\bibnamefont {Zhao}}, \bibinfo
  {author} {\bibfnamefont {B.}~\bibnamefont {Kang}}, \bibinfo {author}
  {\bibfnamefont {J.}~\bibnamefont {Zhang}},\ and\ \bibinfo {author}
  {\bibfnamefont {W.}~\bibnamefont {Ren}},\ }\bibfield  {title} {\bibinfo
  {title} {{Temperature induced Spin Switching in {SmFeO}$_3$ Single
  Crystal}},\ }\href {https://doi.org/10.1038/srep05960} {\bibfield  {journal}
  {\bibinfo  {journal} {Scientific Reports}\ }\textbf {\bibinfo {volume} {4}},\
  \bibinfo {pages} {5960} (\bibinfo {year} {2014})}\BibitemShut {NoStop}%
\bibitem [{\citenamefont {Bertaut}\ \emph {et~al.}(1967)\citenamefont
  {Bertaut}, \citenamefont {Chappert}, \citenamefont {Mareschal}, \citenamefont
  {Rebouillat},\ and\ \citenamefont {Sivardière}}]{Bertaut1967}%
  \BibitemOpen
  \bibfield  {author} {\bibinfo {author} {\bibfnamefont {E.}~\bibnamefont
  {Bertaut}}, \bibinfo {author} {\bibfnamefont {J.}~\bibnamefont {Chappert}},
  \bibinfo {author} {\bibfnamefont {J.}~\bibnamefont {Mareschal}}, \bibinfo
  {author} {\bibfnamefont {J.}~\bibnamefont {Rebouillat}},\ and\ \bibinfo
  {author} {\bibfnamefont {J.}~\bibnamefont {Sivardière}},\ }\bibfield
  {title} {\bibinfo {title} {{Structures magnetiques de TbFeO$_3$}},\ }\href
  {https://doi.org/https://doi.org/10.1016/0038-1098(67)90276-1} {\bibfield
  {journal} {\bibinfo  {journal} {Solid State Communications}\ }\textbf
  {\bibinfo {volume} {5}},\ \bibinfo {pages} {293} (\bibinfo {year}
  {1967})}\BibitemShut {NoStop}%
\bibitem [{\citenamefont {Goodenough}(1963)}]{Goodenough1963}%
  \BibitemOpen
  \bibfield  {author} {\bibinfo {author} {\bibfnamefont {J.}~\bibnamefont
  {Goodenough}},\ }\href@noop {} {\bibinfo {title} {Magnetism and the chemical
  bond}} (\bibinfo {year} {1963})\BibitemShut {NoStop}%
\bibitem [{\citenamefont {Kanamori}(1959)}]{Kanamori1959}%
  \BibitemOpen
  \bibfield  {author} {\bibinfo {author} {\bibfnamefont {J.}~\bibnamefont
  {Kanamori}},\ }\bibfield  {title} {\bibinfo {title} {Superexchange
  interaction and symmetry properties of electron orbitals},\ }\href
  {https://doi.org/https://doi.org/10.1016/0022-3697(59)90061-7} {\bibfield
  {journal} {\bibinfo  {journal} {Journal of Physics and Chemistry of Solids}\
  }\textbf {\bibinfo {volume} {10}},\ \bibinfo {pages} {87 } (\bibinfo {year}
  {1959})}\BibitemShut {NoStop}%
\bibitem [{\citenamefont {Anderson}(1959)}]{Anderson1959}%
  \BibitemOpen
  \bibfield  {author} {\bibinfo {author} {\bibfnamefont {P.~W.}\ \bibnamefont
  {Anderson}},\ }\bibfield  {title} {\bibinfo {title} {New approach to the
  theory of superexchange interactions},\ }\href
  {https://doi.org/10.1103/PhysRev.115.2} {\bibfield  {journal} {\bibinfo
  {journal} {Phys. Rev.}\ }\textbf {\bibinfo {volume} {115}},\ \bibinfo {pages}
  {2} (\bibinfo {year} {1959})}\BibitemShut {NoStop}%
\bibitem [{\citenamefont {Shin}\ and\ \citenamefont
  {Rondinelli}(2020)}]{Shin/Rondinelli2020}%
  \BibitemOpen
  \bibfield  {author} {\bibinfo {author} {\bibfnamefont {Y.}~\bibnamefont
  {Shin}}\ and\ \bibinfo {author} {\bibfnamefont {J.~M.}\ \bibnamefont
  {Rondinelli}},\ }\bibfield  {title} {\bibinfo {title} {Pressure effects on
  magnetism in {Ca}$_2${Mn}$_2${O}$_5$-type ferrites and manganites},\ }\href
  {https://doi.org/10.1103/PhysRevB.102.104426} {\bibfield  {journal} {\bibinfo
   {journal} {Physical Review B}\ }\textbf {\bibinfo {volume} {102}},\ \bibinfo
  {pages} {1} (\bibinfo {year} {2020})}\BibitemShut {NoStop}%
\bibitem [{\citenamefont {Shin}\ and\ \citenamefont
  {Rondinelli}(2022)}]{Shin/Rondinelli2022}%
  \BibitemOpen
  \bibfield  {author} {\bibinfo {author} {\bibfnamefont {Y.}~\bibnamefont
  {Shin}}\ and\ \bibinfo {author} {\bibfnamefont {J.~M.}\ \bibnamefont
  {Rondinelli}},\ }\bibfield  {title} {\bibinfo {title} {Magnetic structure of
  oxygen-deficient perovskite nickelates with ordered vacancies},\ }\href
  {https://doi.org/10.1103/PhysRevResearch.4.L022069} {\bibfield  {journal}
  {\bibinfo  {journal} {Phys. Rev. Res.}\ }\textbf {\bibinfo {volume} {4}},\
  \bibinfo {pages} {L022069} (\bibinfo {year} {2022})}\BibitemShut {NoStop}%
\bibitem [{\citenamefont {Spaldin}(2012)}]{Spaldin2012}%
  \BibitemOpen
  \bibfield  {author} {\bibinfo {author} {\bibfnamefont {N.~A.}\ \bibnamefont
  {Spaldin}},\ }\bibfield  {title} {\bibinfo {title} {A beginner's guide to the
  modern theory of polarization},\ }\href
  {https://doi.org/https://doi.org/10.1016/j.jssc.2012.05.010} {\bibfield
  {journal} {\bibinfo  {journal} {Journal of Solid State Chemistry}\ }\textbf
  {\bibinfo {volume} {195}},\ \bibinfo {pages} {2} (\bibinfo {year}
  {2012})}\BibitemShut {NoStop}%
\bibitem [{\citenamefont {Resta}\ and\ \citenamefont
  {Vanderbilt}(2007)}]{Resta/Vanderbilt2007}%
  \BibitemOpen
  \bibfield  {author} {\bibinfo {author} {\bibfnamefont {R.}~\bibnamefont
  {Resta}}\ and\ \bibinfo {author} {\bibfnamefont {D.}~\bibnamefont
  {Vanderbilt}},\ }\bibinfo {title} {Theory of polarization: A modern
  approach},\ in\ \href {https://doi.org/10.1007/978-3-540-34591-6_2} {\emph
  {\bibinfo {booktitle} {Physics of Ferroelectrics: A Modern Perspective}}}\
  (\bibinfo  {publisher} {Springer Berlin Heidelberg},\ \bibinfo {address}
  {Berlin, Heidelberg},\ \bibinfo {year} {2007})\BibitemShut {NoStop}%
\bibitem [{\citenamefont {Yuk}\ \emph {et~al.}(2017)\citenamefont {Yuk},
  \citenamefont {Pitike}, \citenamefont {Nakhmanson}, \citenamefont
  {Eisenbach}, \citenamefont {Li},\ and\ \citenamefont {Cooper}}]{Yuk2017}%
  \BibitemOpen
  \bibfield  {author} {\bibinfo {author} {\bibfnamefont {S.~F.}\ \bibnamefont
  {Yuk}}, \bibinfo {author} {\bibfnamefont {K.~C.}\ \bibnamefont {Pitike}},
  \bibinfo {author} {\bibfnamefont {S.~M.}\ \bibnamefont {Nakhmanson}},
  \bibinfo {author} {\bibfnamefont {M.}~\bibnamefont {Eisenbach}}, \bibinfo
  {author} {\bibfnamefont {Y.~W.}\ \bibnamefont {Li}},\ and\ \bibinfo {author}
  {\bibfnamefont {V.~R.}\ \bibnamefont {Cooper}},\ }\bibfield  {title}
  {\bibinfo {title} {Towards an accurate description of perovskite
  ferroelectrics: exchange and correlation effects},\ }\href
  {https://doi.org/10.1038/srep43482} {\bibfield  {journal} {\bibinfo
  {journal} {Scientific Reports}\ }\textbf {\bibinfo {volume} {7}},\ \bibinfo
  {pages} {43482} (\bibinfo {year} {2017})}\BibitemShut {NoStop}%
\bibitem [{\citenamefont {Waroquiers}\ \emph {et~al.}(2017)\citenamefont
  {Waroquiers}, \citenamefont {Gonze}, \citenamefont {Rignanese}, \citenamefont
  {Welker-Nieuwoudt}, \citenamefont {Rosowski}, \citenamefont {Göbel},
  \citenamefont {Schenk}, \citenamefont {Degelmann}, \citenamefont {André},
  \citenamefont {Glaum},\ and\ \citenamefont {Hautier}}]{Hautier2017}%
  \BibitemOpen
  \bibfield  {author} {\bibinfo {author} {\bibfnamefont {D.}~\bibnamefont
  {Waroquiers}}, \bibinfo {author} {\bibfnamefont {X.}~\bibnamefont {Gonze}},
  \bibinfo {author} {\bibfnamefont {G.-M.}\ \bibnamefont {Rignanese}}, \bibinfo
  {author} {\bibfnamefont {C.}~\bibnamefont {Welker-Nieuwoudt}}, \bibinfo
  {author} {\bibfnamefont {F.}~\bibnamefont {Rosowski}}, \bibinfo {author}
  {\bibfnamefont {M.}~\bibnamefont {Göbel}}, \bibinfo {author} {\bibfnamefont
  {S.}~\bibnamefont {Schenk}}, \bibinfo {author} {\bibfnamefont
  {P.}~\bibnamefont {Degelmann}}, \bibinfo {author} {\bibfnamefont
  {R.}~\bibnamefont {André}}, \bibinfo {author} {\bibfnamefont
  {R.}~\bibnamefont {Glaum}},\ and\ \bibinfo {author} {\bibfnamefont
  {G.}~\bibnamefont {Hautier}},\ }\bibfield  {title} {\bibinfo {title}
  {Statistical {Analysis} of {Coordination} {Environments} in {Oxides}},\
  }\href {https://doi.org/10.1021/acs.chemmater.7b02766} {\bibfield  {journal}
  {\bibinfo  {journal} {Chemistry of Materials}\ }\textbf {\bibinfo {volume}
  {29}},\ \bibinfo {pages} {8346} (\bibinfo {year} {2017})}\BibitemShut
  {NoStop}%
\bibitem [{\citenamefont {Mishra}\ \emph {et~al.}(2014)\citenamefont {Mishra},
  \citenamefont {Kim}, \citenamefont {Salafranca}, \citenamefont {Kim},
  \citenamefont {Chang}, \citenamefont {Bhattacharya}, \citenamefont {Fong},
  \citenamefont {Pennycook}, \citenamefont {Pantelides},\ and\ \citenamefont
  {Borisevich}}]{Mishra2014}%
  \BibitemOpen
  \bibfield  {author} {\bibinfo {author} {\bibfnamefont {R.}~\bibnamefont
  {Mishra}}, \bibinfo {author} {\bibfnamefont {Y.-M.}\ \bibnamefont {Kim}},
  \bibinfo {author} {\bibfnamefont {J.}~\bibnamefont {Salafranca}}, \bibinfo
  {author} {\bibfnamefont {S.~K.}\ \bibnamefont {Kim}}, \bibinfo {author}
  {\bibfnamefont {S.~H.}\ \bibnamefont {Chang}}, \bibinfo {author}
  {\bibfnamefont {A.}~\bibnamefont {Bhattacharya}}, \bibinfo {author}
  {\bibfnamefont {D.~D.}\ \bibnamefont {Fong}}, \bibinfo {author}
  {\bibfnamefont {S.~J.}\ \bibnamefont {Pennycook}}, \bibinfo {author}
  {\bibfnamefont {S.~T.}\ \bibnamefont {Pantelides}},\ and\ \bibinfo {author}
  {\bibfnamefont {A.~Y.}\ \bibnamefont {Borisevich}},\ }\bibfield  {title}
  {\bibinfo {title} {Oxygen-vacancy-induced polar behavior in
  {(LaFeO$_3$)$_2$/(SrFeO$_3$)} superlattices},\ }\href
  {https://doi.org/10.1021/nl500601d} {\bibfield  {journal} {\bibinfo
  {journal} {Nano Letters}\ }\textbf {\bibinfo {volume} {14}},\ \bibinfo
  {pages} {2694} (\bibinfo {year} {2014})}\BibitemShut {NoStop}%
\bibitem [{\citenamefont {Benedek}(2014)}]{DJ:Benedek2014}%
  \BibitemOpen
  \bibfield  {author} {\bibinfo {author} {\bibfnamefont {N.~A.}\ \bibnamefont
  {Benedek}},\ }\bibfield  {title} {\bibinfo {title} {Origin of
  ferroelectricity in a family of polar oxides: The {Dion}—{Jacobson}
  phases},\ }\href {https://doi.org/10.1021/ic500106a} {\bibfield  {journal}
  {\bibinfo  {journal} {Inorganic Chemistry}\ }\textbf {\bibinfo {volume}
  {53}},\ \bibinfo {pages} {3769} (\bibinfo {year} {2014})}\BibitemShut
  {NoStop}%
\bibitem [{\citenamefont {Lu}\ and\ \citenamefont
  {Rondinelli}(2016)}]{Lu/Rondinelli2016}%
  \BibitemOpen
  \bibfield  {author} {\bibinfo {author} {\bibfnamefont {X.~Z.}\ \bibnamefont
  {Lu}}\ and\ \bibinfo {author} {\bibfnamefont {J.~M.}\ \bibnamefont
  {Rondinelli}},\ }\bibfield  {title} {\bibinfo {title}
  {Epitaxial-strain-induced polar-to-nonpolar transitions in layered oxides},\
  }\href {https://doi.org/10.1038/nmat4664} {\bibfield  {journal} {\bibinfo
  {journal} {Nature Materials}\ }\textbf {\bibinfo {volume} {15}},\ \bibinfo
  {pages} {951} (\bibinfo {year} {2016})}\BibitemShut {NoStop}%
\bibitem [{\citenamefont {Zhang}\ \emph {et~al.}(2019)\citenamefont {Zhang},
  \citenamefont {Solanki}, \citenamefont {Parida}, \citenamefont {Giovanni},
  \citenamefont {Li}, \citenamefont {Jansen}, \citenamefont {Pshenichnikov},\
  and\ \citenamefont {Sum}}]{Zhang/Chien2019}%
  \BibitemOpen
  \bibfield  {author} {\bibinfo {author} {\bibfnamefont {Q.}~\bibnamefont
  {Zhang}}, \bibinfo {author} {\bibfnamefont {A.}~\bibnamefont {Solanki}},
  \bibinfo {author} {\bibfnamefont {K.}~\bibnamefont {Parida}}, \bibinfo
  {author} {\bibfnamefont {D.}~\bibnamefont {Giovanni}}, \bibinfo {author}
  {\bibfnamefont {M.}~\bibnamefont {Li}}, \bibinfo {author} {\bibfnamefont
  {T.~L.~C.}\ \bibnamefont {Jansen}}, \bibinfo {author} {\bibfnamefont {M.~S.}\
  \bibnamefont {Pshenichnikov}},\ and\ \bibinfo {author} {\bibfnamefont
  {T.~C.}\ \bibnamefont {Sum}},\ }\bibfield  {title} {\bibinfo {title} {Tunable
  ferroelectricity in {Ruddlesden}–{Popper} halide perovskites},\ }\href
  {https://doi.org/10.1021/acsami.8b21579} {\bibfield  {journal} {\bibinfo
  {journal} {ACS Applied Materials \& Interfaces}\ }\textbf {\bibinfo {volume}
  {11}},\ \bibinfo {pages} {13523} (\bibinfo {year} {2019})}\BibitemShut
  {NoStop}%
\bibitem [{\citenamefont {Becher}\ \emph {et~al.}(2015)\citenamefont {Becher},
  \citenamefont {Maurel}, \citenamefont {Aschauer}, \citenamefont {Lilienblum},
  \citenamefont {Magén}, \citenamefont {Meier}, \citenamefont {Langenberg},
  \citenamefont {Trassin}, \citenamefont {Blasco}, \citenamefont {Krug},
  \citenamefont {Algarabel}, \citenamefont {Spaldin}, \citenamefont {Pardo},\
  and\ \citenamefont {Fiebig}}]{Becher/Manfred2015}%
  \BibitemOpen
  \bibfield  {author} {\bibinfo {author} {\bibfnamefont {C.}~\bibnamefont
  {Becher}}, \bibinfo {author} {\bibfnamefont {L.}~\bibnamefont {Maurel}},
  \bibinfo {author} {\bibfnamefont {U.}~\bibnamefont {Aschauer}}, \bibinfo
  {author} {\bibfnamefont {M.}~\bibnamefont {Lilienblum}}, \bibinfo {author}
  {\bibfnamefont {C.}~\bibnamefont {Magén}}, \bibinfo {author} {\bibfnamefont
  {D.}~\bibnamefont {Meier}}, \bibinfo {author} {\bibfnamefont
  {E.}~\bibnamefont {Langenberg}}, \bibinfo {author} {\bibfnamefont
  {M.}~\bibnamefont {Trassin}}, \bibinfo {author} {\bibfnamefont
  {J.}~\bibnamefont {Blasco}}, \bibinfo {author} {\bibfnamefont {I.~P.}\
  \bibnamefont {Krug}}, \bibinfo {author} {\bibfnamefont {P.~A.}\ \bibnamefont
  {Algarabel}}, \bibinfo {author} {\bibfnamefont {N.~A.}\ \bibnamefont
  {Spaldin}}, \bibinfo {author} {\bibfnamefont {J.~A.}\ \bibnamefont {Pardo}},\
  and\ \bibinfo {author} {\bibfnamefont {M.}~\bibnamefont {Fiebig}},\
  }\bibfield  {title} {\bibinfo {title} {Strain-induced coupling of electrical
  polarization and structural defects in {SrMnO}$_3$ films},\ }\href
  {https://doi.org/10.1038/nnano.2015.108} {\bibfield  {journal} {\bibinfo
  {journal} {Nature Nanotechnology}\ }\textbf {\bibinfo {volume} {10}},\
  \bibinfo {pages} {661} (\bibinfo {year} {2015})}\BibitemShut {NoStop}%
\bibitem [{\citenamefont {Li}\ \emph {et~al.}(2022)\citenamefont {Li},
  \citenamefont {Yang}, \citenamefont {Deng}, \citenamefont {Zhang},
  \citenamefont {Cheng}, \citenamefont {Guo}, \citenamefont {Zhu},
  \citenamefont {Wang}, \citenamefont {Wang}, \citenamefont {Wu}, \citenamefont
  {Gao}, \citenamefont {Xiang}, \citenamefont {Xing},\ and\ \citenamefont
  {Chen}}]{Li/Chen2022}%
  \BibitemOpen
  \bibfield  {author} {\bibinfo {author} {\bibfnamefont {H.}~\bibnamefont
  {Li}}, \bibinfo {author} {\bibfnamefont {Y.}~\bibnamefont {Yang}}, \bibinfo
  {author} {\bibfnamefont {S.}~\bibnamefont {Deng}}, \bibinfo {author}
  {\bibfnamefont {L.}~\bibnamefont {Zhang}}, \bibinfo {author} {\bibfnamefont
  {S.}~\bibnamefont {Cheng}}, \bibinfo {author} {\bibfnamefont {E.-J.}\
  \bibnamefont {Guo}}, \bibinfo {author} {\bibfnamefont {T.}~\bibnamefont
  {Zhu}}, \bibinfo {author} {\bibfnamefont {H.}~\bibnamefont {Wang}}, \bibinfo
  {author} {\bibfnamefont {J.}~\bibnamefont {Wang}}, \bibinfo {author}
  {\bibfnamefont {M.}~\bibnamefont {Wu}}, \bibinfo {author} {\bibfnamefont
  {P.}~\bibnamefont {Gao}}, \bibinfo {author} {\bibfnamefont {H.}~\bibnamefont
  {Xiang}}, \bibinfo {author} {\bibfnamefont {X.}~\bibnamefont {Xing}},\ and\
  \bibinfo {author} {\bibfnamefont {J.}~\bibnamefont {Chen}},\ }\bibfield
  {title} {\bibinfo {title} {{Role of oxygen vacancies in colossal polarization
  in SmFeO$_{3-\delta}$ thin films}},\ }\href
  {https://doi.org/10.1126/sciadv.abm8550} {\bibfield  {journal} {\bibinfo
  {journal} {Science Advances}\ }\textbf {\bibinfo {volume} {8}},\ \bibinfo
  {pages} {eabm8550} (\bibinfo {year} {2022})}\BibitemShut {NoStop}%
\bibitem [{\citenamefont {Kahlenberg}\ \emph {et~al.}(2000)\citenamefont
  {Kahlenberg}, \citenamefont {Fischer},\ and\ \citenamefont
  {Shaw}}]{Kahlenberg2000}%
  \BibitemOpen
  \bibfield  {author} {\bibinfo {author} {\bibfnamefont {V.}~\bibnamefont
  {Kahlenberg}}, \bibinfo {author} {\bibfnamefont {R.~X.}\ \bibnamefont
  {Fischer}},\ and\ \bibinfo {author} {\bibfnamefont {C.~S.}\ \bibnamefont
  {Shaw}},\ }\bibfield  {title} {\bibinfo {title} {Rietveld analysis of
  dicalcium aluminate ({Ca$_2$Al$_2$O$_5$}) – a new high pressure phase with
  the brownmillerite-type structure},\ }\href
  {https://doi.org/doi:10.2138/am-2000-0722} {\bibfield  {journal} {\bibinfo
  {journal} {American Mineralogist}\ }\textbf {\bibinfo {volume} {85}},\
  \bibinfo {pages} {1061} (\bibinfo {year} {2000})}\BibitemShut {NoStop}%
\bibitem [{\citenamefont {Von~Schenck}\ and\ \citenamefont
  {Müller-Buschbaum}(1973)}]{Sr2In2O5}%
  \BibitemOpen
  \bibfield  {author} {\bibinfo {author} {\bibfnamefont {R.}~\bibnamefont
  {Von~Schenck}}\ and\ \bibinfo {author} {\bibfnamefont {H.}~\bibnamefont
  {Müller-Buschbaum}},\ }\bibfield  {title} {\bibinfo {title} {{Über ein
  neues Erdalkalimetall-Oxoindat: Sr$_2$In$_2$O$_5$}},\ }\href
  {https://doi.org/https://doi.org/10.1002/zaac.19733950216} {\bibfield
  {journal} {\bibinfo  {journal} {Zeitschrift für anorganische und allgemeine
  Chemie}\ }\textbf {\bibinfo {volume} {395}},\ \bibinfo {pages} {280}
  (\bibinfo {year} {1973})}\BibitemShut {NoStop}%
\bibitem [{\citenamefont {Perdew}\ \emph {et~al.}(1996)\citenamefont {Perdew},
  \citenamefont {Burke},\ and\ \citenamefont {Ernzerhof}}]{PBE}%
  \BibitemOpen
  \bibfield  {author} {\bibinfo {author} {\bibfnamefont {J.~P.}\ \bibnamefont
  {Perdew}}, \bibinfo {author} {\bibfnamefont {K.}~\bibnamefont {Burke}},\ and\
  \bibinfo {author} {\bibfnamefont {M.}~\bibnamefont {Ernzerhof}},\ }\bibfield
  {title} {\bibinfo {title} {Generalized gradient approximation made simple},\
  }\href {https://doi.org/10.1103/PhysRevLett.77.3865} {\bibfield  {journal}
  {\bibinfo  {journal} {Phys. Rev. Lett.}\ }\textbf {\bibinfo {volume} {77}},\
  \bibinfo {pages} {3865} (\bibinfo {year} {1996})}\BibitemShut {NoStop}%
\bibitem [{\citenamefont {Giannozzi}\ \emph {et~al.}(2009)\citenamefont
  {Giannozzi}, \citenamefont {Baroni}, \citenamefont {Bonini}, \citenamefont
  {Calandra}, \citenamefont {Car}, \citenamefont {Cavazzoni}, \citenamefont
  {Ceresoli}, \citenamefont {Chiarotti}, \citenamefont {Cococcioni},
  \citenamefont {Dabo}, \citenamefont {Corso}, \citenamefont {de~Gironcoli},
  \citenamefont {Fabris}, \citenamefont {Fratesi}, \citenamefont {Gebauer},
  \citenamefont {Gerstmann}, \citenamefont {Gougoussis}, \citenamefont
  {Kokalj}, \citenamefont {Lazzeri}, \citenamefont {Martin-Samos},
  \citenamefont {Marzari}, \citenamefont {Mauri}, \citenamefont {Mazzarello},
  \citenamefont {Paolini}, \citenamefont {Pasquarello}, \citenamefont
  {Paulatto}, \citenamefont {Sbraccia}, \citenamefont {Scandolo}, \citenamefont
  {Sclauzero}, \citenamefont {Seitsonen}, \citenamefont {Smogunov},
  \citenamefont {Umari},\ and\ \citenamefont {Wentzcovitch}}]{Giannozzi_2009}%
  \BibitemOpen
  \bibfield  {author} {\bibinfo {author} {\bibfnamefont {P.}~\bibnamefont
  {Giannozzi}}, \bibinfo {author} {\bibfnamefont {S.}~\bibnamefont {Baroni}},
  \bibinfo {author} {\bibfnamefont {N.}~\bibnamefont {Bonini}}, \bibinfo
  {author} {\bibfnamefont {M.}~\bibnamefont {Calandra}}, \bibinfo {author}
  {\bibfnamefont {R.}~\bibnamefont {Car}}, \bibinfo {author} {\bibfnamefont
  {C.}~\bibnamefont {Cavazzoni}}, \bibinfo {author} {\bibfnamefont
  {D.}~\bibnamefont {Ceresoli}}, \bibinfo {author} {\bibfnamefont {G.~L.}\
  \bibnamefont {Chiarotti}}, \bibinfo {author} {\bibfnamefont {M.}~\bibnamefont
  {Cococcioni}}, \bibinfo {author} {\bibfnamefont {I.}~\bibnamefont {Dabo}},
  \bibinfo {author} {\bibfnamefont {A.~D.}\ \bibnamefont {Corso}}, \bibinfo
  {author} {\bibfnamefont {S.}~\bibnamefont {de~Gironcoli}}, \bibinfo {author}
  {\bibfnamefont {S.}~\bibnamefont {Fabris}}, \bibinfo {author} {\bibfnamefont
  {G.}~\bibnamefont {Fratesi}}, \bibinfo {author} {\bibfnamefont
  {R.}~\bibnamefont {Gebauer}}, \bibinfo {author} {\bibfnamefont
  {U.}~\bibnamefont {Gerstmann}}, \bibinfo {author} {\bibfnamefont
  {C.}~\bibnamefont {Gougoussis}}, \bibinfo {author} {\bibfnamefont
  {A.}~\bibnamefont {Kokalj}}, \bibinfo {author} {\bibfnamefont
  {M.}~\bibnamefont {Lazzeri}}, \bibinfo {author} {\bibfnamefont
  {L.}~\bibnamefont {Martin-Samos}}, \bibinfo {author} {\bibfnamefont
  {N.}~\bibnamefont {Marzari}}, \bibinfo {author} {\bibfnamefont
  {F.}~\bibnamefont {Mauri}}, \bibinfo {author} {\bibfnamefont
  {R.}~\bibnamefont {Mazzarello}}, \bibinfo {author} {\bibfnamefont
  {S.}~\bibnamefont {Paolini}}, \bibinfo {author} {\bibfnamefont
  {A.}~\bibnamefont {Pasquarello}}, \bibinfo {author} {\bibfnamefont
  {L.}~\bibnamefont {Paulatto}}, \bibinfo {author} {\bibfnamefont
  {C.}~\bibnamefont {Sbraccia}}, \bibinfo {author} {\bibfnamefont
  {S.}~\bibnamefont {Scandolo}}, \bibinfo {author} {\bibfnamefont
  {G.}~\bibnamefont {Sclauzero}}, \bibinfo {author} {\bibfnamefont {A.~P.}\
  \bibnamefont {Seitsonen}}, \bibinfo {author} {\bibfnamefont {A.}~\bibnamefont
  {Smogunov}}, \bibinfo {author} {\bibfnamefont {P.}~\bibnamefont {Umari}},\
  and\ \bibinfo {author} {\bibfnamefont {R.~M.}\ \bibnamefont {Wentzcovitch}},\
  }\bibfield  {title} {\bibinfo {title} {{QUANTUM} {ESPRESSO}: a modular and
  open-source software project for quantum simulations of materials},\ }\href
  {https://doi.org/10.1088/0953-8984/21/39/395502} {\bibfield  {journal}
  {\bibinfo  {journal} {Journal of Physics: Condensed Matter}\ }\textbf
  {\bibinfo {volume} {21}},\ \bibinfo {pages} {395502} (\bibinfo {year}
  {2009})}\BibitemShut {NoStop}%
\bibitem [{\citenamefont {Giannozzi}\ \emph {et~al.}(2017)\citenamefont
  {Giannozzi}, \citenamefont {Andreussi}, \citenamefont {Brumme}, \citenamefont
  {Bunau}, \citenamefont {Nardelli}, \citenamefont {Calandra}, \citenamefont
  {Car}, \citenamefont {Cavazzoni}, \citenamefont {Ceresoli}, \citenamefont
  {Cococcioni}, \citenamefont {Colonna}, \citenamefont {Carnimeo},
  \citenamefont {Corso}, \citenamefont {de~Gironcoli}, \citenamefont {Delugas},
  \citenamefont {DiStasio}, \citenamefont {Ferretti}, \citenamefont {Floris},
  \citenamefont {Fratesi}, \citenamefont {Fugallo}, \citenamefont {Gebauer},
  \citenamefont {Gerstmann}, \citenamefont {Giustino}, \citenamefont {Gorni},
  \citenamefont {Jia}, \citenamefont {Kawamura}, \citenamefont {Ko},
  \citenamefont {Kokalj}, \citenamefont {Kü{\c{c}}ükbenli}, \citenamefont
  {Lazzeri}, \citenamefont {Marsili}, \citenamefont {Marzari}, \citenamefont
  {Mauri}, \citenamefont {Nguyen}, \citenamefont {Nguyen}, \citenamefont {de-la
  Roza}, \citenamefont {Paulatto}, \citenamefont {Ponc{\'{e}}}, \citenamefont
  {Rocca}, \citenamefont {Sabatini}, \citenamefont {Santra}, \citenamefont
  {Schlipf}, \citenamefont {Seitsonen}, \citenamefont {Smogunov}, \citenamefont
  {Timrov}, \citenamefont {Thonhauser}, \citenamefont {Umari}, \citenamefont
  {Vast}, \citenamefont {Wu},\ and\ \citenamefont {Baroni}}]{Giannozzi_2017}%
  \BibitemOpen
  \bibfield  {author} {\bibinfo {author} {\bibfnamefont {P.}~\bibnamefont
  {Giannozzi}}, \bibinfo {author} {\bibfnamefont {O.}~\bibnamefont
  {Andreussi}}, \bibinfo {author} {\bibfnamefont {T.}~\bibnamefont {Brumme}},
  \bibinfo {author} {\bibfnamefont {O.}~\bibnamefont {Bunau}}, \bibinfo
  {author} {\bibfnamefont {M.~B.}\ \bibnamefont {Nardelli}}, \bibinfo {author}
  {\bibfnamefont {M.}~\bibnamefont {Calandra}}, \bibinfo {author}
  {\bibfnamefont {R.}~\bibnamefont {Car}}, \bibinfo {author} {\bibfnamefont
  {C.}~\bibnamefont {Cavazzoni}}, \bibinfo {author} {\bibfnamefont
  {D.}~\bibnamefont {Ceresoli}}, \bibinfo {author} {\bibfnamefont
  {M.}~\bibnamefont {Cococcioni}}, \bibinfo {author} {\bibfnamefont
  {N.}~\bibnamefont {Colonna}}, \bibinfo {author} {\bibfnamefont
  {I.}~\bibnamefont {Carnimeo}}, \bibinfo {author} {\bibfnamefont {A.~D.}\
  \bibnamefont {Corso}}, \bibinfo {author} {\bibfnamefont {S.}~\bibnamefont
  {de~Gironcoli}}, \bibinfo {author} {\bibfnamefont {P.}~\bibnamefont
  {Delugas}}, \bibinfo {author} {\bibfnamefont {R.~A.}\ \bibnamefont
  {DiStasio}}, \bibinfo {author} {\bibfnamefont {A.}~\bibnamefont {Ferretti}},
  \bibinfo {author} {\bibfnamefont {A.}~\bibnamefont {Floris}}, \bibinfo
  {author} {\bibfnamefont {G.}~\bibnamefont {Fratesi}}, \bibinfo {author}
  {\bibfnamefont {G.}~\bibnamefont {Fugallo}}, \bibinfo {author} {\bibfnamefont
  {R.}~\bibnamefont {Gebauer}}, \bibinfo {author} {\bibfnamefont
  {U.}~\bibnamefont {Gerstmann}}, \bibinfo {author} {\bibfnamefont
  {F.}~\bibnamefont {Giustino}}, \bibinfo {author} {\bibfnamefont
  {T.}~\bibnamefont {Gorni}}, \bibinfo {author} {\bibfnamefont
  {J.}~\bibnamefont {Jia}}, \bibinfo {author} {\bibfnamefont {M.}~\bibnamefont
  {Kawamura}}, \bibinfo {author} {\bibfnamefont {H.-Y.}\ \bibnamefont {Ko}},
  \bibinfo {author} {\bibfnamefont {A.}~\bibnamefont {Kokalj}}, \bibinfo
  {author} {\bibfnamefont {E.}~\bibnamefont {Kü{\c{c}}ükbenli}}, \bibinfo
  {author} {\bibfnamefont {M.}~\bibnamefont {Lazzeri}}, \bibinfo {author}
  {\bibfnamefont {M.}~\bibnamefont {Marsili}}, \bibinfo {author} {\bibfnamefont
  {N.}~\bibnamefont {Marzari}}, \bibinfo {author} {\bibfnamefont
  {F.}~\bibnamefont {Mauri}}, \bibinfo {author} {\bibfnamefont {N.~L.}\
  \bibnamefont {Nguyen}}, \bibinfo {author} {\bibfnamefont {H.-V.}\
  \bibnamefont {Nguyen}}, \bibinfo {author} {\bibfnamefont {A.~O.}\
  \bibnamefont {de-la Roza}}, \bibinfo {author} {\bibfnamefont
  {L.}~\bibnamefont {Paulatto}}, \bibinfo {author} {\bibfnamefont
  {S.}~\bibnamefont {Ponc{\'{e}}}}, \bibinfo {author} {\bibfnamefont
  {D.}~\bibnamefont {Rocca}}, \bibinfo {author} {\bibfnamefont
  {R.}~\bibnamefont {Sabatini}}, \bibinfo {author} {\bibfnamefont
  {B.}~\bibnamefont {Santra}}, \bibinfo {author} {\bibfnamefont
  {M.}~\bibnamefont {Schlipf}}, \bibinfo {author} {\bibfnamefont {A.~P.}\
  \bibnamefont {Seitsonen}}, \bibinfo {author} {\bibfnamefont {A.}~\bibnamefont
  {Smogunov}}, \bibinfo {author} {\bibfnamefont {I.}~\bibnamefont {Timrov}},
  \bibinfo {author} {\bibfnamefont {T.}~\bibnamefont {Thonhauser}}, \bibinfo
  {author} {\bibfnamefont {P.}~\bibnamefont {Umari}}, \bibinfo {author}
  {\bibfnamefont {N.}~\bibnamefont {Vast}}, \bibinfo {author} {\bibfnamefont
  {X.}~\bibnamefont {Wu}},\ and\ \bibinfo {author} {\bibfnamefont
  {S.}~\bibnamefont {Baroni}},\ }\bibfield  {title} {\bibinfo {title} {Advanced
  capabilities for materials modelling with quantum {ESPRESSO}},\ }\href
  {https://doi.org/10.1088/1361-648x/aa8f79} {\bibfield  {journal} {\bibinfo
  {journal} {Journal of Physics: Condensed Matter}\ }\textbf {\bibinfo {volume}
  {29}},\ \bibinfo {pages} {465901} (\bibinfo {year} {2017})}\BibitemShut
  {NoStop}%
\bibitem [{\citenamefont {Giannozzi}\ \emph {et~al.}(2020)\citenamefont
  {Giannozzi}, \citenamefont {Baseggio}, \citenamefont {Bonfà}, \citenamefont
  {Brunato}, \citenamefont {Car}, \citenamefont {Carnimeo}, \citenamefont
  {Cavazzoni}, \citenamefont {de~Gironcoli}, \citenamefont {Delugas},
  \citenamefont {Ferrari~Ruffino}, \citenamefont {Ferretti}, \citenamefont
  {Marzari}, \citenamefont {Timrov}, \citenamefont {Urru},\ and\ \citenamefont
  {Baroni}}]{Giannozzi_2020}%
  \BibitemOpen
  \bibfield  {author} {\bibinfo {author} {\bibfnamefont {P.}~\bibnamefont
  {Giannozzi}}, \bibinfo {author} {\bibfnamefont {O.}~\bibnamefont {Baseggio}},
  \bibinfo {author} {\bibfnamefont {P.}~\bibnamefont {Bonfà}}, \bibinfo
  {author} {\bibfnamefont {D.}~\bibnamefont {Brunato}}, \bibinfo {author}
  {\bibfnamefont {R.}~\bibnamefont {Car}}, \bibinfo {author} {\bibfnamefont
  {I.}~\bibnamefont {Carnimeo}}, \bibinfo {author} {\bibfnamefont
  {C.}~\bibnamefont {Cavazzoni}}, \bibinfo {author} {\bibfnamefont
  {S.}~\bibnamefont {de~Gironcoli}}, \bibinfo {author} {\bibfnamefont
  {P.}~\bibnamefont {Delugas}}, \bibinfo {author} {\bibfnamefont
  {F.}~\bibnamefont {Ferrari~Ruffino}}, \bibinfo {author} {\bibfnamefont
  {A.}~\bibnamefont {Ferretti}}, \bibinfo {author} {\bibfnamefont
  {N.}~\bibnamefont {Marzari}}, \bibinfo {author} {\bibfnamefont
  {I.}~\bibnamefont {Timrov}}, \bibinfo {author} {\bibfnamefont
  {A.}~\bibnamefont {Urru}},\ and\ \bibinfo {author} {\bibfnamefont
  {S.}~\bibnamefont {Baroni}},\ }\bibfield  {title} {\bibinfo {title} {Quantum
  espresso toward the exascale},\ }\href {https://doi.org/10.1063/5.0005082}
  {\bibfield  {journal} {\bibinfo  {journal} {The Journal of Chemical Physics}\
  }\textbf {\bibinfo {volume} {152}},\ \bibinfo {pages} {154105} (\bibinfo
  {year} {2020})}\BibitemShut {NoStop}%
\bibitem [{\citenamefont {Cococcioni}\ and\ \citenamefont
  {de~Gironcoli}(2005)}]{Cococcioni2005}%
  \BibitemOpen
  \bibfield  {author} {\bibinfo {author} {\bibfnamefont {M.}~\bibnamefont
  {Cococcioni}}\ and\ \bibinfo {author} {\bibfnamefont {S.}~\bibnamefont
  {de~Gironcoli}},\ }\bibfield  {title} {\bibinfo {title} {Linear response
  approach to the calculation of the effective interaction parameters in the
  $\mathrm{LDA}+\mathrm{U}$ method},\ }\href
  {https://doi.org/10.1103/PhysRevB.71.035105} {\bibfield  {journal} {\bibinfo
  {journal} {Phys. Rev. B}\ }\textbf {\bibinfo {volume} {71}},\ \bibinfo
  {pages} {035105} (\bibinfo {year} {2005})}\BibitemShut {NoStop}%
\bibitem [{\citenamefont {Galakhov}\ \emph {et~al.}(2010)\citenamefont
  {Galakhov}, \citenamefont {Kurmaev}, \citenamefont {Kuepper}, \citenamefont
  {Neumann}, \citenamefont {McLeod}, \citenamefont {Moewes}, \citenamefont
  {Leonidov},\ and\ \citenamefont {Kozhevnikov}}]{Galakhov/Kozhevnikov2010}%
  \BibitemOpen
  \bibfield  {author} {\bibinfo {author} {\bibfnamefont {V.~R.}\ \bibnamefont
  {Galakhov}}, \bibinfo {author} {\bibfnamefont {E.~Z.}\ \bibnamefont
  {Kurmaev}}, \bibinfo {author} {\bibfnamefont {K.}~\bibnamefont {Kuepper}},
  \bibinfo {author} {\bibfnamefont {M.}~\bibnamefont {Neumann}}, \bibinfo
  {author} {\bibfnamefont {J.~A.}\ \bibnamefont {McLeod}}, \bibinfo {author}
  {\bibfnamefont {A.}~\bibnamefont {Moewes}}, \bibinfo {author} {\bibfnamefont
  {I.~A.}\ \bibnamefont {Leonidov}},\ and\ \bibinfo {author} {\bibfnamefont
  {V.~L.}\ \bibnamefont {Kozhevnikov}},\ }\bibfield  {title} {\bibinfo {title}
  {Valence band structure and x-ray spectra of oxygen-deficient ferrites
  srfeox},\ }\href {https://doi.org/10.1021/jp909091s} {\bibfield  {journal}
  {\bibinfo  {journal} {The Journal of Physical Chemistry C}\ }\textbf
  {\bibinfo {volume} {114}},\ \bibinfo {pages} {5154} (\bibinfo {year}
  {2010})}\BibitemShut {NoStop}%
\bibitem [{\citenamefont {{Dal Corso}}(2014)}]{DalCorso2014}%
  \BibitemOpen
  \bibfield  {author} {\bibinfo {author} {\bibfnamefont {A.}~\bibnamefont {{Dal
  Corso}}},\ }\bibfield  {title} {\bibinfo {title} {{Pseudopotentials periodic
  table: From H to Pu}},\ }\href
  {https://doi.org/https://doi.org/10.1016/j.commatsci.2014.07.043} {\bibfield
  {journal} {\bibinfo  {journal} {Computational Materials Science}\ }\textbf
  {\bibinfo {volume} {95}},\ \bibinfo {pages} {337} (\bibinfo {year}
  {2014})}\BibitemShut {NoStop}%
\bibitem [{\citenamefont {Topsakal}\ and\ \citenamefont
  {Wentzcovitch}(2014)}]{Topsakal2014}%
  \BibitemOpen
  \bibfield  {author} {\bibinfo {author} {\bibfnamefont {M.}~\bibnamefont
  {Topsakal}}\ and\ \bibinfo {author} {\bibfnamefont {R.}~\bibnamefont
  {Wentzcovitch}},\ }\bibfield  {title} {\bibinfo {title} {Accurate projected
  augmented wave ({PAW}) datasets for rare-earth elements ({RE}={La}–{Lu})},\
  }\href {https://doi.org/https://doi.org/10.1016/j.commatsci.2014.07.030}
  {\bibfield  {journal} {\bibinfo  {journal} {Computational Materials Science}\
  }\textbf {\bibinfo {volume} {95}},\ \bibinfo {pages} {263} (\bibinfo {year}
  {2014})}\BibitemShut {NoStop}%
\bibitem [{\citenamefont {Shannon}(1976)}]{Shannon1976}%
  \BibitemOpen
  \bibfield  {author} {\bibinfo {author} {\bibfnamefont {R.~D.}\ \bibnamefont
  {Shannon}},\ }\bibfield  {title} {\bibinfo {title} {{Revised effective ionic
  radii and systematic studies of interatomic distances in halides and
  chalcogenides}},\ }\href {https://doi.org/10.1107/S0567739476001551}
  {\bibfield  {journal} {\bibinfo  {journal} {Acta Crystallographica Section
  A}\ }\textbf {\bibinfo {volume} {32}},\ \bibinfo {pages} {751} (\bibinfo
  {year} {1976})}\BibitemShut {NoStop}%
\bibitem [{\citenamefont {Wagner}\ \emph {et~al.}(2018)\citenamefont {Wagner},
  \citenamefont {Puggioni},\ and\ \citenamefont
  {Rondinelli}}]{Wagner/Rondinelli2018}%
  \BibitemOpen
  \bibfield  {author} {\bibinfo {author} {\bibfnamefont {N.}~\bibnamefont
  {Wagner}}, \bibinfo {author} {\bibfnamefont {D.}~\bibnamefont {Puggioni}},\
  and\ \bibinfo {author} {\bibfnamefont {J.~M.}\ \bibnamefont {Rondinelli}},\
  }\bibfield  {title} {\bibinfo {title} {Learning from correlations based on
  local structure: Rare-earth nickelates revisited},\ }\href
  {https://doi.org/10.1021/acs.jcim.8b00411} {\bibfield  {journal} {\bibinfo
  {journal} {Journal of Chemical Information and Modeling}\ }\textbf {\bibinfo
  {volume} {58}},\ \bibinfo {pages} {2491} (\bibinfo {year}
  {2018})}\BibitemShut {NoStop}%
\bibitem [{\citenamefont {Guo}\ \emph {et~al.}(2018)\citenamefont {Guo},
  \citenamefont {Li}, \citenamefont {Zhao}, \citenamefont {Hu}, \citenamefont
  {Chang}, \citenamefont {Kuo}, \citenamefont {Schmidt}, \citenamefont
  {Piovano}, \citenamefont {Pi}, \citenamefont {Sobolev}, \citenamefont
  {Khomskii}, \citenamefont {Tjeng},\ and\ \citenamefont
  {Komarek}}]{Guo/Khomskii2018}%
  \BibitemOpen
  \bibfield  {author} {\bibinfo {author} {\bibfnamefont {H.}~\bibnamefont
  {Guo}}, \bibinfo {author} {\bibfnamefont {Z.~W.}\ \bibnamefont {Li}},
  \bibinfo {author} {\bibfnamefont {L.}~\bibnamefont {Zhao}}, \bibinfo {author}
  {\bibfnamefont {Z.}~\bibnamefont {Hu}}, \bibinfo {author} {\bibfnamefont
  {C.~F.}\ \bibnamefont {Chang}}, \bibinfo {author} {\bibfnamefont {C.-Y.}\
  \bibnamefont {Kuo}}, \bibinfo {author} {\bibfnamefont {W.}~\bibnamefont
  {Schmidt}}, \bibinfo {author} {\bibfnamefont {A.}~\bibnamefont {Piovano}},
  \bibinfo {author} {\bibfnamefont {T.~W.}\ \bibnamefont {Pi}}, \bibinfo
  {author} {\bibfnamefont {O.}~\bibnamefont {Sobolev}}, \bibinfo {author}
  {\bibfnamefont {D.~I.}\ \bibnamefont {Khomskii}}, \bibinfo {author}
  {\bibfnamefont {L.~H.}\ \bibnamefont {Tjeng}},\ and\ \bibinfo {author}
  {\bibfnamefont {A.~C.}\ \bibnamefont {Komarek}},\ }\bibfield  {title}
  {\bibinfo {title} {{Antiferromagnetic correlations in the metallic strongly
  correlated transition metal oxide LaNiO$_3$}},\ }\href
  {https://doi.org/10.1038/s41467-017-02524-x} {\bibfield  {journal} {\bibinfo
  {journal} {Nature Communications}\ }\textbf {\bibinfo {volume} {9}},\
  \bibinfo {pages} {43} (\bibinfo {year} {2018})}\BibitemShut {NoStop}%
\bibitem [{\citenamefont {Momma}\ and\ \citenamefont {Izumi}(2011)}]{VESTA}%
  \BibitemOpen
  \bibfield  {author} {\bibinfo {author} {\bibfnamefont {K.}~\bibnamefont
  {Momma}}\ and\ \bibinfo {author} {\bibfnamefont {F.}~\bibnamefont {Izumi}},\
  }\bibfield  {title} {\bibinfo {title} {{{\it VESTA3} for three-dimensional
  visualization of crystal, volumetric and morphology data}},\ }\href
  {https://doi.org/10.1107/S0021889811038970} {\bibfield  {journal} {\bibinfo
  {journal} {Journal of Applied Crystallography}\ }\textbf {\bibinfo {volume}
  {44}},\ \bibinfo {pages} {1272} (\bibinfo {year} {2011})}\BibitemShut
  {NoStop}%
\bibitem [{\citenamefont {Govoni}\ \emph {et~al.}(2019)\citenamefont {Govoni},
  \citenamefont {Munakami}, \citenamefont {Tanikanti}, \citenamefont {Skone},
  \citenamefont {Runesha}, \citenamefont {Giberti}, \citenamefont {de~Pablo},\
  and\ \citenamefont {Galli}}]{Qresp}%
  \BibitemOpen
  \bibfield  {author} {\bibinfo {author} {\bibfnamefont {M.}~\bibnamefont
  {Govoni}}, \bibinfo {author} {\bibfnamefont {M.}~\bibnamefont {Munakami}},
  \bibinfo {author} {\bibfnamefont {A.}~\bibnamefont {Tanikanti}}, \bibinfo
  {author} {\bibfnamefont {J.~H.}\ \bibnamefont {Skone}}, \bibinfo {author}
  {\bibfnamefont {H.~B.}\ \bibnamefont {Runesha}}, \bibinfo {author}
  {\bibfnamefont {F.}~\bibnamefont {Giberti}}, \bibinfo {author} {\bibfnamefont
  {J.}~\bibnamefont {de~Pablo}},\ and\ \bibinfo {author} {\bibfnamefont
  {G.}~\bibnamefont {Galli}},\ }\bibfield  {title} {\bibinfo {title} {Qresp, a
  tool for curating, discovering and exploring reproducible scientific
  papers},\ }\href {https://doi.org/10.1038/sdata.2019.2} {\bibfield  {journal}
  {\bibinfo  {journal} {Scientific Data}\ }\textbf {\bibinfo {volume} {6}},\
  \bibinfo {pages} {190002} (\bibinfo {year} {2019})}\BibitemShut {NoStop}%
\end{thebibliography}

%

\end{document}


\noindent
{\bf Supplementary Information for}\\
\vspace{-0.3 cm}

\noindent{\bf Tunable ferroelectricity in oxygen-deficient perovskites}
\vspace{0.5cm}

\noindent
Yongjin Shin$^1$ and Giulia Galli$^{1,2,3,*}$\\
$^1$\textit{\footnotesize Pritzker School of Molecular Engineering, University of Chicago, Chicago, Illinois 60637, United States} \vspace{-0.25 cm}\\
$^2$\textit{\footnotesize Department of Chemistry, University of Chicago, Chicago, Illinois 60637, United States} \vspace{-0.25 cm} \\
$^3$\textit{\footnotesize Center for Molecular Engineering and Materials Science Division, Argonne National Laboratory, Lemont, Illinois 60439, United States} \vspace{-0.25 cm} \\
$^*${\small electronic mail: gagalli@uchicago.edu}\\


\def\degree{$^\circ$\xspace}
\newcommand\Tstrut{\rule{0pt}{2.6ex}}         
\newcommand\Bstrut{\rule[-0.9ex]{0pt}{0pt}}   
\newcommand\RAFO{$R_{1/3}A_{2/3}$FeO$_{2.67}$\xspace}
\newcommand\NCFO{Nd$_{1/3}$Ca$_{2/3}$FeO$_{2.67}$\xspace}
\newcommand\LSFO{La$_{1/3}$Sr$_{2/3}$FeO$_{2.67}$\xspace}
\newcommand\SFO{SrFeO$_{2.5}$\xspace}
\newcommand\CFO{CaFeO$_{2.5}$\xspace}
\newcommand\BFO{BaFeO$_{2.5}$\xspace}
\newcommand\AFO{$A$FeO$_{2.5}$\xspace}
\newcommand\Ccm{$\mu\mathrm{C/cm}^2$\xspace}

\renewcommand{\baselinestretch}{0.75}\normalsize
{
\hypersetup{linkcolor=black}
\tableofcontents
}
\renewcommand{\baselinestretch}{1.25}\normalsize

\newpage

\section{Effect of Hubbard $U$ on computed properties}

\begin{figure}[h]
\centering
\includegraphics[width=0.95\columnwidth]{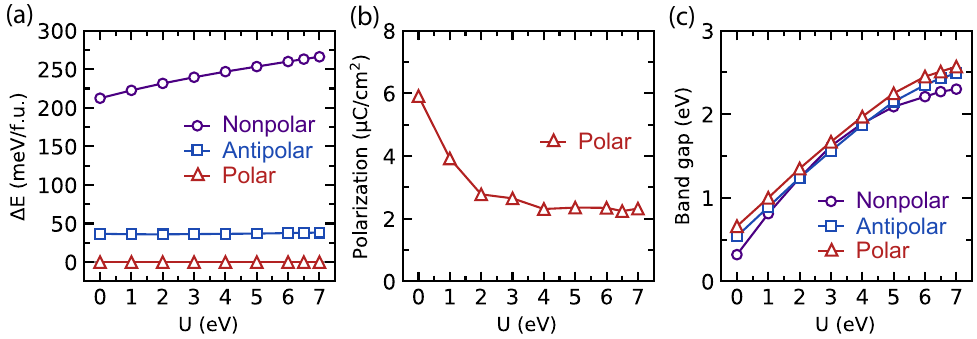}
\caption{
(a) Energy difference ($\Delta$E) between nonpolar, polar and antipolar phases of \NCFO, as a function of the Hubbard $U$ term used in DFT total energy calculations. (b) Value of the polarization as a function of $U$. (b) Fundamental band gap as a function of $U$.
}
\label{fig:U_test}
\end{figure}

In \autoref{fig:U_test}, we show how several computed properties of the \NCFO Grenier structure vary as a function of the value of $U$ used in our DFT calculations.
We find that these properties are not sensitive to the $U$ value varying from 0 to 7 eV.
Irrespective of the value of $U$, we find that the ground state of \NCFO is the polar phase, and that
the energy difference between polar and antipolar phases varies within a small energy range, from 36.0 to 38.2 meV/f.u.
On the other hand, the energy difference between nonpolar and polar phases increases from 212 to 266 meV/f.u. as $U$ increases.

The spontaneous polarization decreases and converges to a finite value, as $U$ increases, as shown in \autoref{fig:U_test}(b).
The polarization is 5.91 \Ccm at $U$ = 0 eV,
 decreases to 2.7 \Ccm at $U$ = 2 eV, and  converges to $\sim$2.3 \Ccm for $U$ $>$ 4 eV.
The same convergence to a finite value of the polarization as a function of $U$ is also found in hybrid improper ferroelectric Ca$_3$Ti$_2$O$_7$ \cite{Li/Liu2018}.
%
%
Using a large $U$ value leads to an increase in localization of the $d$-electronic orbitals  of the transition metal, which in turns reduces the polarization of ferroelectric structures arising from the second-order Jahn-Teller effect \cite{Din2020}.
However, the hybrid improper ferroelectricity and the ferroelectricity found in Grenier phases rely on the rotation of polyhedral units without requiring any off-centering of the transition metal, and thus the value of the spontaneous polarization is insensitive to the choice of the $U$ value.

Finally, in \autoref{fig:U_test}(c) we show that the fundamental band gap of \NCFO increases as a function of $U$ for polar, antipolar, and nonpolar phases.
\NCFO remains insulating for all values of the Hubbard $U$, including 0 eV.
%
Our calculations are consistent with the general effect of $U$ \cite{Cococcioni2005,Shin/Rondinelli2020} on the computed value of the band gap observed in other materials,  where $U$ lowers (increases) the energy of occupied (empty) $d$ bands, thereby increasing the band gap.

\section{Effect of spin-orbit coupling on computed properties}

We investigated the effect of spin-orbit coupling (SOC) on the ferroelectric properties  of \NCFO Grenier structure.
In calculations including SOC, we utilized optimized norm-conserving Vanderbilt (ONCV) pseudopotentials,
a kinetic energy cutoff of 75 Ry, a $k$-mesh of $3\times8\times8$, and a Hubbard $U$ value of 6.5 eV for the structural relaxation.
We used the Berry phase method to evaluate the spontaneous polarization.
We considered  G-AFM order with a collinear spin moment along the $z$-direction (polar direction) in the $Pmc2_1$ phase, based on experimental observations \cite{Shimakawa/Martin2019}.

We find that the energy difference between the nonpolar and the polar phase is similar in calculations with and without SOC.
Without SOC, the energy difference is 261.6 meV/f.u.; when including SOC the difference slightly reduces to   260.0 meV/f.u.
In addition, we find that
the spontaneous polarization is relatively insensitive to the inclusion of SOC in our calculations: it increases from 1.71 \Ccm to 1.73 \Ccm when including SOC. We conclude that overall, the impact of SOC on the ferroelectric properties of the Grenier structure of \NCFO is not significant.

\section{Electronic and magnetic properties of the \NCFO Grenier structure}

\begin{figure}[h]
\centering
\includegraphics[width=0.80\columnwidth]{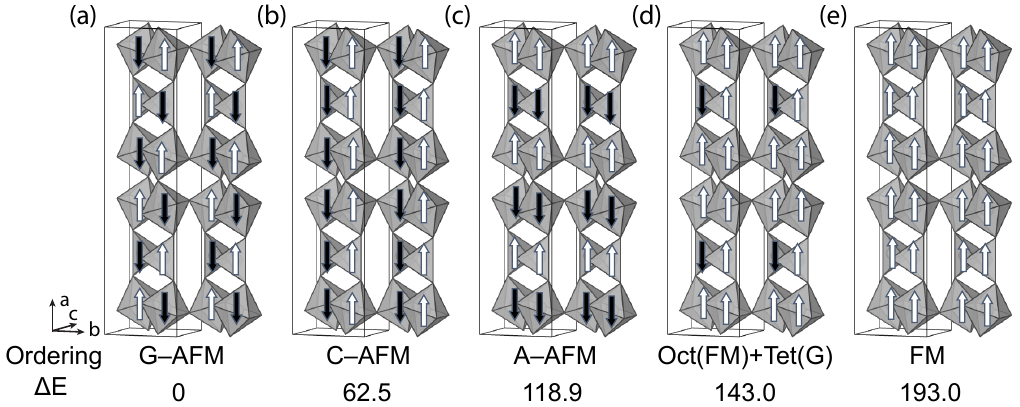}
\caption{
Representation of the structure of \NCFO with different magnetic ordering and their energy differences (meV/f.u.) compared with the ground state magnetic order (G-AFM). Nd, Ca, and O atoms are omitted for clarity.
G-AFM corresponds to a checkerboard arrangement of spin moments, while C-AFM involves columnar ordering of spin moments along the $a$-direction. A-AFM refers to a layered arrangement of spins with the same orientation. In Oct(FM)+Tet(G), ferromagnetic coupling exists within octahedral layers, whereas antiferromagnetic coupling occurs in tetrahedral layers. FM denotes a configuration where all spin moments align in the same direction.
}
\label{fig:S_MagOrder}
\end{figure}

When studying the \NCFO Grenier structure, we investigated various magnetic ordering commonly observed in perovskite oxides, including ferromagnetic (FM), G-type  (G-AFM), A-type (A-AFM), and C-type antiferromagnetic (C-AFM). We also considered the magnetic order found in the LaCoO$_{2.67}$ Grenier structure \cite{Biskup2014}, where the octahedral layers exhibit FM-coupling within the  octahedral layers (with charge ordering and breathing distortion), and AFM-copuling within the tetrahedral layers as shown in \autoref{fig:S_MagOrder}(d).
We found  that the ground state magnetic order of \NCFO is G-AFM, which is consistent with the  Goodenough-Kanamori-Anderson rule. The Fe$^{3+}$ ions with a high-spin $d^5$ configuration favors AFM-coupling along the Fe--O--Fe path. 
The second most stable magnetic order is C-AFM, and its energy difference with respect to G-AFM is 62.5 meV/f.u.
Based on this analysis,
we considred a G-AFM order for all  \RAFO systems in our study.

\begin{figure}[h]
\centering
\includegraphics[width=0.85\columnwidth]{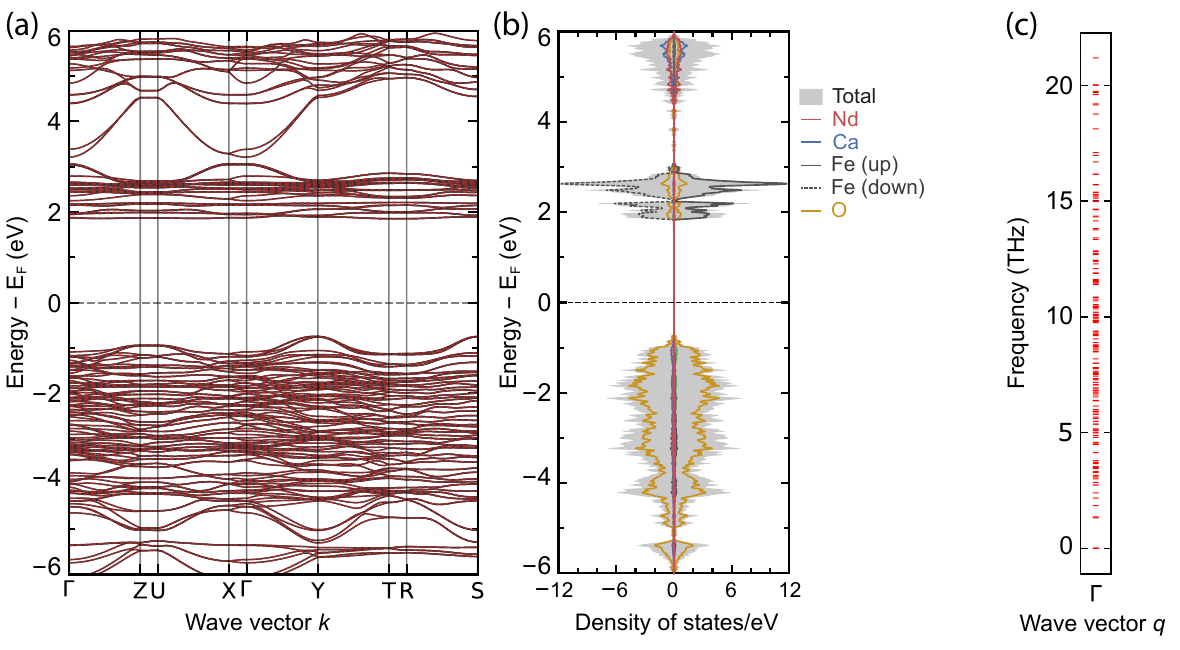}
\caption{
Electronic structure of the polar Grenier phase of  \NCFO with $Pmc2_1$ space group and G-type antiferromagnetic order. (a) Band diagram with up- and down-spin components drawn as solid black and dashed darkred lines, respectively. (b) Projected density of states of constituent elements.
(c) Phonon frequencies at the $\Gamma$ point ($6a_p \times 2\sqrt{2}a_p \times 2\sqrt{2}a_p$).
}
\label{fig:S_band}
\end{figure}

Next we computed the ground state properties of the polar Grenier phase of \NCFO with  space group $Pmc2_1$.
The electronic band structure and density of states are shown in \autoref{fig:S_band}(a) and (b).
The conduction band is primarily composed of Fe-$d$ orbitals, while the valence band is composed of O-$p$ orbitals. The band gap is 2.51 eV when we apply a $U$ value of 6.5 eV.
Due to the antiferromagnetic (AFM) order of the system,
the spin-up and spin-down components of Fe-orbitals are symmetric in the band structure and density of states.

To investigate the thermodynamic stability of the polar phase, we performed a frozen phonon calculation using the phonopy package \cite{Phonopy}. As shown in \autoref{fig:S_band}(c), no imaginary phonon frequencies were found at the $\Gamma$ point, indicating a stable  polar phase.

The crystallographic information of the polar \NCFO phase as obtained from total energy minimizations is provided in \autoref{table:NCFO}, where we also give  the Born effective charges (BEC) of each Wyckoff positions, obtained from our phonon calculations. Due to the high computational cost of BEC calculations, we used an energy cutoff of 50 Ry and 1$\times$3$\times$3 $k$-grid.

\begin{table}[h]
\caption{\label{table:NCFO} 
Crystallographic data of \NCFO with $Pmc2_1$ space group, as obtained in our total energy optimization, with $U$ = 6.5 eV. Lattice parameters are $a$ = 11.26434 \AA, $b$ = 5.57643 \AA, and $c$ = 5.60952 \AA. $Z^*_{xx}$, $Z^*_{yy}$, and $Z^*_{zz}$ are Born effective charge of each ion along $a$, $b$, and $c$-directions, respectively.
}
\centering
\begin{ruledtabular}
\begin{tabular}{l|cccc|ccc}
Atom & Site & x & y & z & $Z^*_{xx}$ & $Z^*_{yy}$ & $Z^*_{zz}$\\
\hline\Tstrut
Nd1  & 2a  &  0.00000 &  0.23455  & 0.81368  & 3.491 & 4.281 & 4.161 \\
Ca1  & 4c  &  0.31676 &  0.22965  & 0.71415  & 2.666 & 2.692 & 2.569 \\
Fe1  & 4c  &  0.82598 &  0.73780  & 0.76116  & 3.638 & 3.730 & 3.778 \\
Fe2  & 2b  &  0.50000 &  0.69001  & 0.70412  & 3.824 & 2.525 & 2.891 \\
O1   & 4c  &  0.20388 &  0.45395  & -0.03237 & -1.916 & -2.513 & -2.776 \\
O2   & 4c  &  0.86034 &  -0.03799 & 0.05114  & -2.022 & -2.892 & -2.853 \\
O3   & 2b  &  0.50000 &  0.37880  & 0.86289  & -1.763 & -1.827 & -2.749 \\
O4   & 2a  &  0.00000 &  0.65300  & 0.73409  & -2.628 & -2.410 & -2.058 \\
O5   & 4c  &  0.35708 &  0.83497  & 0.79854  & -3.538 & -2.087 & -1.657 \\
\end{tabular}
\end{ruledtabular}
\end{table}

\section{Effect of cation ordering in \NCFO}

\begin{figure}[h]
\centering
\includegraphics[width=0.90\columnwidth]{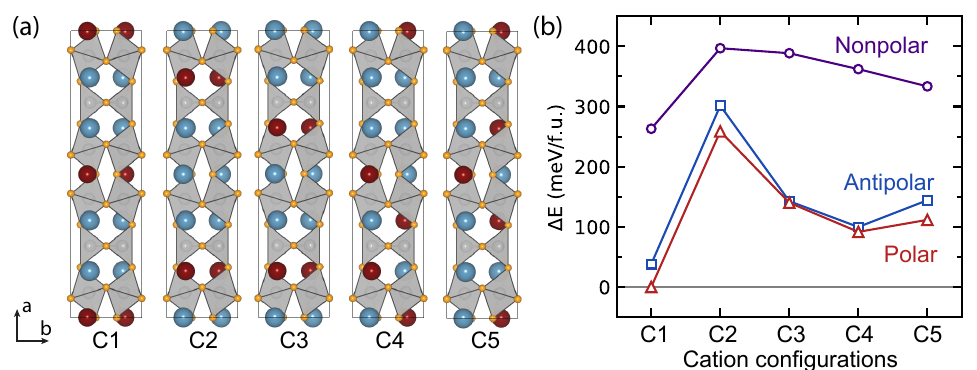}
\caption{
(a) Various cation configurations in the \NCFO Grenier structure (labeled C1 to C5, see text). Red and blue spheres represent Nd and Ca ions, respectively.
(b) Total energy differences s between polar, antipolar, and nonpolar phases as a function of the configurations shown in (a).
}
\label{fig:cation_order}
\end{figure}

We performed calculations for  different geometrical arrangements of Nd ($R$) and Ca ($A$) ions in \NCFO, in order to examine the impact of cation ordering in the Grenier structure.
The configurations of \NCFO considered in this work are shown in \autoref{fig:cation_order}(a) and labeled as C1 to C5.
In C1, the Nd ions occupy the cation sites between the two octahedral layers, and Ca ions occupy the cation sites in neighboring tetrahedral layers.
Because the symmetry of these two sites is different, the space groups of nonpolar, antipolar, and polar phases of C1 are $Pmma$, $Pbcm$, and $Pmc2_1$, respectively, and are equivalent to those of a Grenier structure with homogeneous mixing of $R$- and $A$-ions.
The configurations labeled (C2--5) are constructed to represent  cation mixing different from that of C1  and from homogeneously distributed $R$- and $A$-ions.
The configurations C2 and C3 have layered ordering of cations such that equivalent type of cations are arranged on the same $ab$-plane in the unit cell.
On the other hand, C4 and C5 are devised to have nonequivalent cations on the same $ab$-plane.

\autoref{fig:cation_order}(b) shows the energy differences between polar, nonpolar, and antipolar phases for different configurations (C1--5).
For all configurations, we find that polar phases are the lowest in energy, followed by the antipolar phases, indicating that in Grenier structures, ferroelectricity can be realized with various cation arrangements.
In the main text, we  discuss  the ferroelectricity of the Grenier structure based on our results for the ``C1" configuration, which was found to be the most stable one, in agreement  with the experimental observation that the $R$-ions preferably occupy the sites between two octahedral layers \cite{Hudspeth2009}.
The stability of the C1 configuration can be rationalized as follows: the presence of oxygen vacancies creates extra room for cations adjacent to the tetrahedral layers, compared to cations within the octahedral layers. As a result, C1, having relatively large $A$-cations near the tetrahedral layers and small rare earth ions within the octahedral layers, is energetically favoured.
The energy difference between polar and antipolar phases differ, depending on the configuration chosen: it  is 38.0 meV/f.u. for C1, as small as 2.54 meV/f.u. for C3, and as large as 42.8 meV/f.u. for C2.
The energy difference between antipolar and nonpolar phases is particularly small for C2,  with the total energy of the nonpolar phase only 138.2 meV/f.u. higher than that of the polar phase.
The other configurations exhibit  energy differences between nonpolar and antipolar phases in the range 222 -- 270 meV/f.u.

\section{Energy differences as a function of cation in \RAFO}

\begin{figure}[h]
\centering
\includegraphics[width=0.9\columnwidth]{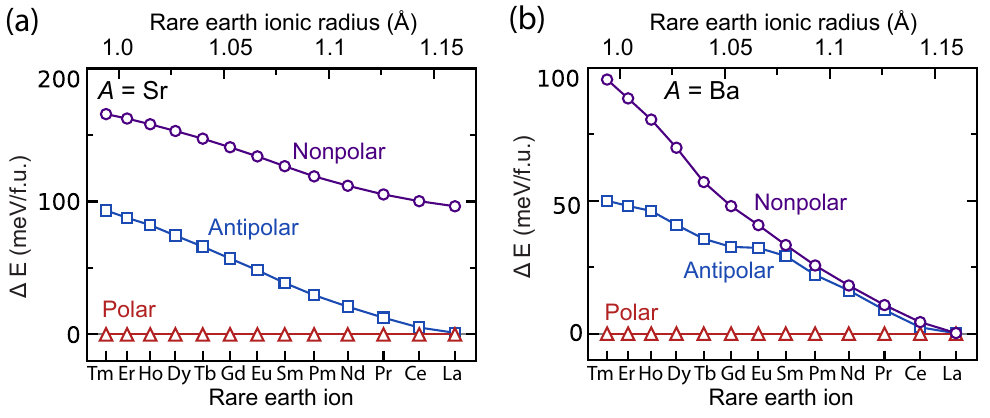}
\caption{
Total energy differences between polar, antipolar and nonpolar phases of the Grenier \RAFO solids, as a function of the rare earth ions. (a) and (b)) show results for $A$ = Sr and $A$ = Ba, respectively.
}
\label{fig:S_energetics}
\end{figure}

In \autoref{fig:S_energetics},
we show the energy differences between the polar, antipolar, and nonpolar phases of the Grenier structures of \RAFO with $A$= Sr or Ba. We note that these plots are equivalent to those appearing in \autoref{fig:cation_effect}(b).
The energy landscape for  $A$ = Sr is similar to that of $A$ = Ca.
The polar phases are more stable than nonpolar and antipolar phases 
and the energy differences gradually increase as the radii of the rare earth elements decrease.
We observe an energy difference of 93.2 meV/f.u. for $R$ = Pm,  decreasing to 1.22 meV/f.u. for $R$ = La.

As mentioned in the main text, the presence of a large $A$ cation significantly reduces the energy difference between polar and nonpolar phases, 
while the effect of the $R$-ion size  on the energy difference is similar for all $A$-cations.
For  $A$ = Ca, the energy difference decreases from 320 to 236 meV/f.u. as the atomic radii of the $R$-ion increases,
while for $A$ = Sr, it decreases from 165 to 96 meV/f.u.
The main structural difference between the nonpolar and the antipolar phases is the presence of twisted tetrahedral units.
The higher energy of the nonpolar phase is associated with an undercoordinated $A$-cation, where the larger ionic radius of Sr compared to Ca reduces the energy difference between nonpolar and other phases.
The size of the $R$-ion does not play a significant role in the energy difference between nonpolar and antipolar phases because
local  atomic structures in proximity of  the $R$ ion are similar.
Indeed, the energy difference between the nonpolar and the antipolar phases is similar for  different $R$-ions.

Also for $A$ = Ba, the polar phase is the ground state for all $R$-ions; however
the energy differences between antipolar and nonpolar phases are significantly smaller compared to the corresponding ones for $A$ = Ca or Sr.
When $R$ is larger than Eu, the nonpolar and antipolar curves in \autoref{fig:S_energetics}(b) nearly overlap.
For ($R$ = La, $A$ = Ba), all distortion patterns fall within 1 meV/f.u.
This small energy difference may be attributed to the small structural differences between the antipolar and nonpolar phases, \textit{e.g.} see the values of $\theta_\textrm{twist}$ in \autoref{fig:S_descriptors}(c).

\section{Effect of cation size on the structural properties of polar phases of \RAFO}

\begin{figure}[h]
\centering
\includegraphics[width=0.43\columnwidth]{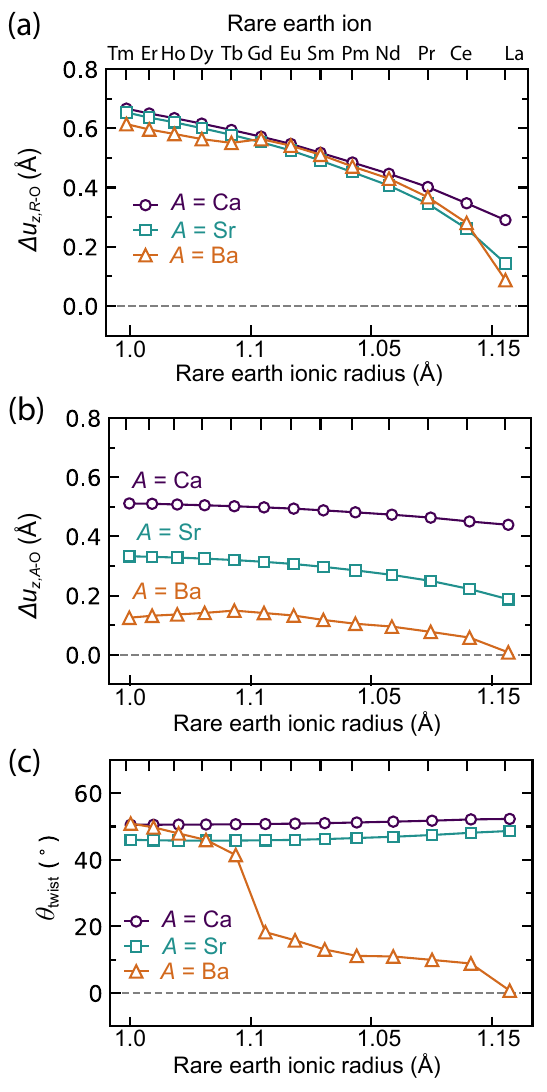}
\caption{
Structural descriptors of the polar \RAFO:
displacements of (a) $R$--O and (b) $A$--O pairs along the polar direction, and (c) twist angle of tetrahedral chains as a function of rare earth elements.
}
\label{fig:S_descriptors}
\end{figure}

In \autoref{fig:S_descriptors},
we present the structural descriptors that characterize the polar phase of the \RAFO Grenier structure. 
These descriptors include the cation-anion displacement along the $z$-direction in the $R$O layer ($\Delta u_{z,R-\rm{O}}$), the displacement in the $A$O layer ($\Delta u_{z,A-\rm{O}}$), and the twist angle of the tetrahedral chains ($\theta_{\rm twist}$).
The values of these descriptors is zero in the nonpolar phase and the spontaneous polarization varies substantially as a function of these descriptors  (\autoref{fig:polarization}).
%
We found a clear correlations between these descriptors and the size of the $R$-ion and the $A$-cations, which we describe in detail below. In this section, we refer to the \RAFO compounds as $(R,A)$ for brevity.

First, we found that the in-plane atomic displacements, $\Delta u_{z,R-\rm{O}}$ and $\Delta u_{z,A-\rm{O}}$, are primarily determined by the size of the $R$-ion and the $A$-cations, respectively [\autoref{fig:S_descriptors}(a) and (b)].
The largest value of $\Delta u_{z,R-\rm{O}}$ is observed for the smallest cations ($R$ = Tm, $A$ = Ca), reaching 0.67 \AA, while the lowest value 0.09 \AA\xspace is observed for the largest cations ($R$ = La, $A$ = Ba).
However, the effect of the $A$-cation on $\Delta u_{z,R-\rm{O}}$ indicates that $\Delta u_{z,R-\rm{O}}$ is mainly determined by the rare earth species.
The curves representing  $\Delta u_{A-\rm{O}}$ as a function of the $A$-cations are well-separated in \autoref{fig:S_descriptors}(c), by approximately 0.2 \AA. In contrast, those as a function of the $R$-ion differ by approximately 0.1 \AA.
Thus the impact of the $R$-ion on the value of $\Delta u_{z,A-\rm{O}}$ is smaller than that of the $A$ on $\Delta u_{A-\rm{O}}$.

To illustrate the effect of $\theta_{\rm twist}$,
we focus first on the $A$ = Ca and Sr systems; as for  $A$ = Ba we find a  more complex behavior that we discuss separately.
Similar to the trend observed for $\Delta u_{z,A-\rm{O}}$, 
we find that a smaller cation (Ca) leads to a larger $\theta_{\rm twist}$.
This trend can be understood by observing that the twisting of the tetrahedral chain reduces the empty space surrounding $A$ by moving oxygen atoms closer to the $A$-cations.
Hence the presence of smaller $A$-cation results in a larger twisting motion. Because Ca is smaller than Sr, ($R$, Ca) solids exhibit larger twist angles than ($R$, Sr).
Indeed, in brownmillerites smaller $A$-cations induce larger twist angles in tetrahedral chains \cite{Parsons2009,Young2015,Tian2018}. Since the distortion patterns found in  Grenier phases are similar to those of browmillerites, it is reasonable to expect that smaller $R$-ions and $A$-cations induce larger distortions also in Grenier phases.

\begin{figure}[h]
\centering
\includegraphics[width=0.85\columnwidth]{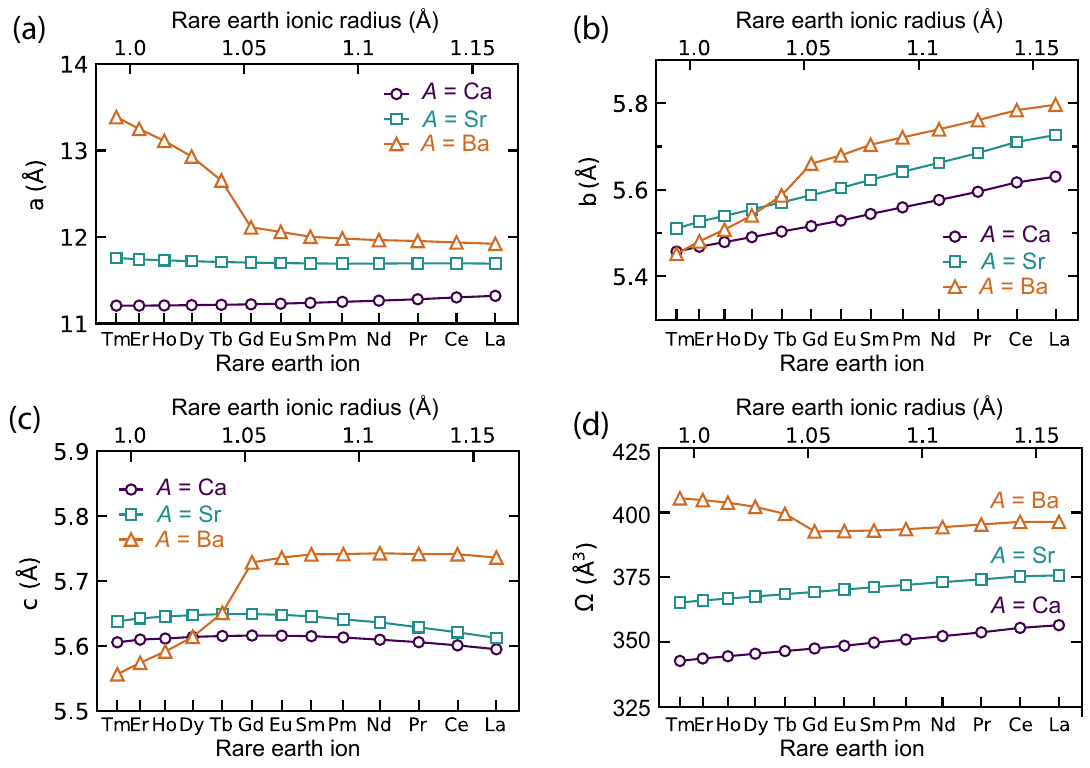}
\caption{
Lattice parameters ($a$, $b$, and $c$) and unit cell volume ($\Omega$) of polar phases of \RAFO as a function of rare earth ion $R$.
}
\label{fig:S_lattice}
\end{figure}

Finally, we explain the unique behavior of $\theta_{\rm twist}$ for $A$ = Ba.
While for $A$ = Ba, $\Delta u_{z,R-\rm{O}}$ and $\Delta u_{z,A-\rm{O}}$ follows the same trend observed for $A$ = Ca and Sr,
the trend of $\theta_{\rm twist}$ is different.
In particular, we find a transition point in going from
($R$ = Tb, $A$ = Ba) to  ($R$ = Gd, $A$ = Ba), where $\theta_{\rm twist}$ drops from 41.5\degree to 18.3\degree.
When examining the atomic structures of the solids with $A$ = Ba,
we find that in the presence of smaller $R$ ions,   the apical Fe--O bonds of the octahedral units are elongated, resulting in structures where the tetrahedral layers are far apart from the octahedral layers.
In such cases, the tetrahedral chains are decoupled from the octahedral units, leading to significantly higher $\theta_{\rm twist}$.
These structural properties are consistent with the lattice parameters and unit cell volumes found for  the polar phases (\autoref{fig:S_lattice}), which clearly points at structural deformations.

It is important to note that such structural deformations occur in the configuration "C1" (see Fig.  \autoref{fig:cation_order}).
%
It is reasonable to expect that instead of elongated bonds in the system, cation mixing may be favored, which would reduce the bond elongation.
Indeed, the incorporation of Ba into  octahedral layers is experimentally observed in $R_{1.2}$Ba$_{1.2}$Ca$_{0.6}$Fe$_3$O$_8$ ($R$ = Gd, Tb) \cite{Shimakawa/Martin2019}.

\section{Structural  distortions in brownmillerite and Grenier structures}

\begin{figure}[h]
\centering
\includegraphics[width=0.6\columnwidth]{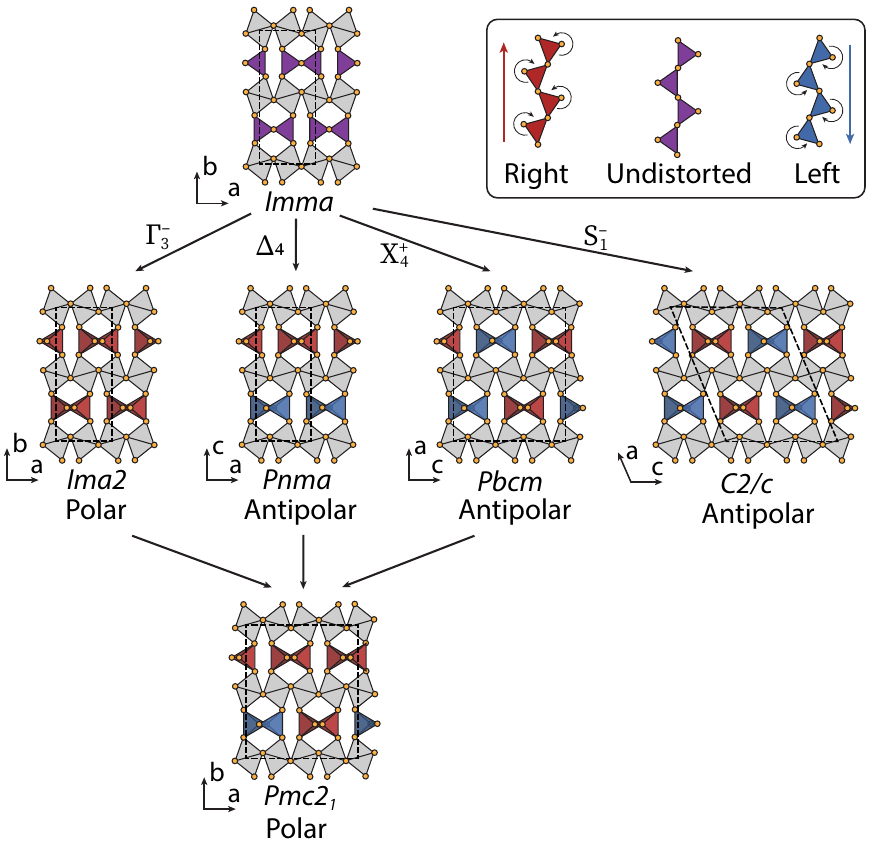}
\caption{Structural distortions leading to symmetry breaking in the brownmillerite structure. Undistorted tetrahedral chains are represented in purple, and right- and left-handed chains are represented in red and blue, respectively. Unit cells are shown with dashed black lines. The $A$-cations are omitted for clarity. Note that the polar structure with the $Ima2$ space group could be equivalently labeled with the $I2bm$ \cite{Young2015} and $I2mb$ \cite{Parsons2009,Luo/Hayward2013} space groups, if different crystallographic axes are chosen. In a similar fashion, $Pbcm$ is equivalent to the $Pcmb$ space group \cite{Parsons2009}.
}
\label{fig:S_BM_map}
\end{figure}

\begin{figure}[h]
\centering
\includegraphics[width=0.6\columnwidth]{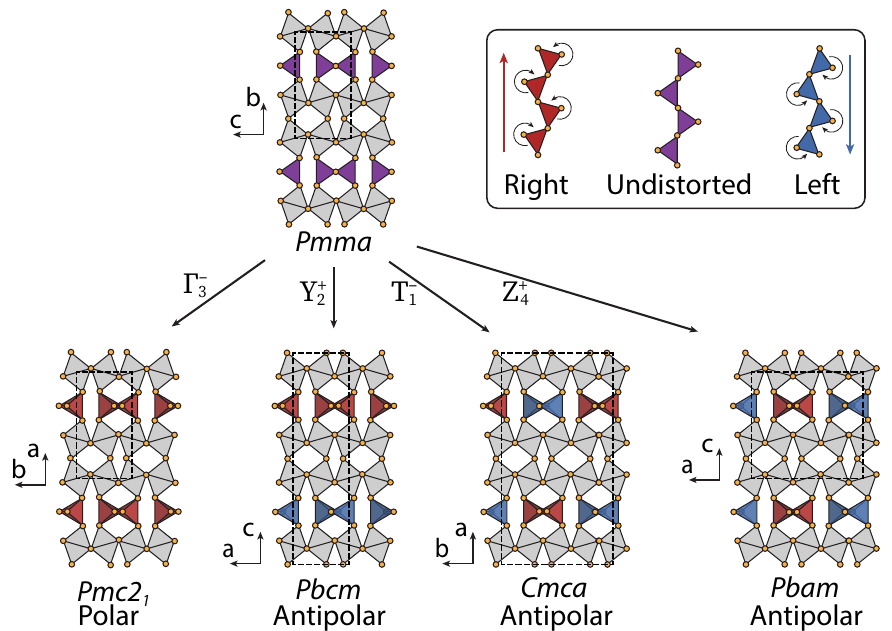}
\caption{
Structural distortions leading to symmetry breaking in the Grenier structure. Undistorted tetrahedral chains are represented in purple, and right- and left-handed chains are represented in red and blue, respectively. Unit cells are shown with dashed black lines. The $R$- and $A$-cations are omitted for clarity. Note that the polar structure with the $Pmc2_1$ space group could be equivalently labeled with the $P2_1ma$ space group, if different crystallographic axes are chosen. In a similar fashion, $Pbcm$ is equivalent to the $Pbma$ space group \cite{Luo/Hayward2013}.
}
\label{fig:S_Grenier_map}
\end{figure}

In \autoref{fig:S_BM_map} and \autoref{fig:S_Grenier_map},
we present structures originating from distortions of the brownmillerite and Grenier phases, respectively, where we indicate the space group of each structure.
It can be challenging to distinguish distortion patterns in these structures, especially if they lead to systems of similar stability. For example, it has been established that the ground state structure of \SFO has $Pbcm$ space group, but due to insufficient resolution in several experiments, different space groups had been previously assigned to this system \cite{Auckett2012}. In general, because both brownmillerite and Grenier structures share similar structural features and show relatively minor variations in oxygen deficiencies, it can be challenging to precisely assign  structures in experiments. By identifying the space group associated with each structure, one can help the analysis of samples and the interpretation of their structural motifs, investigate phase transitions in oxygen-deficient phases, and analyze properties related to symmetry breaking.

In the main text, we mainly discussed the twisting of tetrahedral chains where the chains in the same tetrahedral layers exhibit a consistent twisting direction corresponding to space groups $Imma$, $Ima2$, $Pnma$.
However, we can obtain additional antipolar phases with larger unit cells where the tetrahedral chains with opposite twisting directions alternate within the same layers.
The $Pbcm$ and $C2/c$ space groups both exhibit `intra-planar' alternation of left- and right-handed chains but differ in the stacking of the tetrahedral layers.
In the $C2/c$ space group, the same handed chains are aligned along the diagonal direction ($a$-direction of $C2/c$ in \autoref{fig:S_BM_map}).
In the $Pbcm$ space group, the same handed chains are aligned in a zigzag pattern along the stacking direction ($a$-direction of $Pbcm$ in \autoref{fig:S_BM_map}).
Another type of polar phase, with the $Pmc2_1$ space group, was suggested by Tian \textit{et al.} \cite{Tian2018} for SrCoO$_{2.5}$ brownmillerite, by combining distortions of $Ima2$, $Pnma$, and $Pbcm$.
As a result, the structure with $Pmc2_1$ space group has an unequal number of right- and left-handed chains (three right-handed and one left-handed chains in one unit cell) and a polar phase can be stabilized.

We performed DFT calculations for the $A$FeO$_{2.5}$ brownmillerite structures and we report our results in \autoref{table:Supp_energetics}.
The antipolar $Pnma$ structure is the ground state of \CFO, and the antipolar $C2/c$ structure is the ground state of \SFO and \BFO.
We focus on  the energy difference between the nonpolar and polar phases compared to the ground state antipolar phases.
For \CFO, where the $A$-cation ionic radius is small, the polar phase is 22.24 meV/\CFO higher and the nonpolar phase is 330.1 meV/\CFO higher than the antipolar phase.
For \SFO,
the energy difference between the antipolar and polar phases is instead very small (less than 1 meV/\SFO), but both phases exhibit considerably lower energy compared to that of the nonpolar phase (165 meV/\SFO).
Finally, for $A$ = Ba, we observe a competitive energy landscape among all nonpolar, polar, and antipolar phases (with energy differences less than 1 meV/\BFO).
%


We also conducted a similar analysis for the Grenier structure, as shown in \autoref{fig:S_Grenier_map},
utilizing the ISODISTORT \cite{Campbell:ISODISTORT} and FINDSYM software \cite{FINDSYM}.
By alternating the left- and right-handed chains within the same tetrahedral layer, we identified two antipolar phases with $Cmca$ and $Pbam$ space groups, as illustrated in \autoref{fig:S_Grenier_map}.
These two space groups differ in the stacking arrangement of tetrahedral layers.
In the $Cmca$ structure, left- and right-handed chains alternate along the stacking direction ($a$-direction of the $Cmca$ phase), whereas in the $Pbam$ structure the same type of twisted chains are vertically aligned along the stacking direction ($a$-direction in $Pbam$ phase).%

We compared the stability of each phase in the case of \NCFO using DFT calculations and found that the energies of these phases are considerably higher than that of the $Pmc2_1$ phase.
Specifically, the $Pbcm$ space group structure is 38 meV/f.u. higher than that with the space group $Pmc2_1$, while $Cmca$ and $Pbam$ structures are 79 and 77 meV/f.u. higher than the polar one, respectively.
Therefore, in the main text, we mainly discuss the distortion patterns of the high-symmetry $Pmma$ (nonpolar) structure, and the ground state polar ($Pmc2_1$), and the most stable antipolar ($Pbcm$) phases.

\begin{table}[h]
\caption{\label{table:Supp_energetics} 
Energy differences ($\Delta E$, meV/f.u.) between between polar, nonpolar and antipolar phases relative to their respective ground state (with energy 0) for \AFO brownmillerite structure ($A$ = Ca, Sr, Ba) and the Grenier structure of \NCFO.
}
\centering
\begin{ruledtabular}
\begin{tabular}{l|cccc|l|c}
Space group & CaFeO$_{2.5}$ & SrFeO$_{2.5}$ & BaFeO$_{2.5}$ & & Space group & \NCFO \\
\hline\Tstrut
$Imma$ (non-polar)    & 330.1 & 165.0   & 0.82  & &  $Pmma$ (non-polar)   & 263.1 \\
$Ima2$ (polar)        & 22.24 & 0.98    & 0.79  & &  $Pmc2_1$ (polar)     & 0     \\
$Pnma$ (anti-polar)   & 0     & 2.32    & 0.91  & &  $Pbcm$ (anti-polar)  & 38.01 \\
$Pbcm$ (anti-polar)   & 26.40 & 0.17    & 0.65  & &  $Cmca$ (anti-polar)  & 79.21 \\
$Pmc2_1$ (polar)      & 12.81 & 0.83    & 7.09  & &  $Pbam$ (anti-polar)  & 76.80 \\
$C2/c$ (anti-polar)   & 26.33 & 0       & 0     & &     &    \\
\end{tabular}
\end{ruledtabular}
\end{table}

\newpage


%